\pdfoutput=1
\RequirePackage[2021-06-01]{latexrelease}
\documentclass[]{pnastwo}
\usepackage[center,width=9.125in,height=11.875in,pdftex]{crop}

\url{}
\copyrightyear{2026}
\issuedate{November 25, 2026}
\volume{Volume}
\issuenumber{Issue}
\footlineauthor{N. Innocenti}

\usepackage{url}
\usepackage{booktabs}
\usepackage{tikz}
\usetikzlibrary{positioning,arrows.meta,calc,plotmarks,decorations.pathreplacing,fit}

\definecolor{okBlack}{HTML}{000000}\definecolor{okOrange}{HTML}{E69F00}\definecolor{okSky}{HTML}{56B4E9}
\definecolor{okGreen}{HTML}{009E73}\definecolor{okYellow}{HTML}{F0E442}\definecolor{okBlue}{HTML}{0072B2}
\definecolor{okVermillion}{HTML}{D55E00}\definecolor{okPurple}{HTML}{CC79A7}
\tikzset{every picture/.append style={font=\sffamily\scriptsize, line join=round, line cap=round},
  every node/.append style={text=black!85}}
\newcommand{\tool}[1]{\textsf{#1}}
\newcommand{\mm}{\tool{mawmetal}}
\newcommand{\Sig}{\Sigma}
\newcommand{\none}{\varnothing}

\makeatletter
\newcounter{ntable}
\newcommand{\tabcaption}[2]{\refstepcounter{ntable}\savelabel{#1}%
  {\fignumfont Table~\thentable.\figtextfont\hskip8\p@\relax #2\par}\vskip4\p@}
\renewcommand\subsection{\@startsection{subsection}{2}{\z@}{10\p@\@plus 1ex \@minus .2ex}%
  {-0.6em \@plus .2em}{\normalfont\large\bfseries\subsectionfont}}
\AtBeginDocument{\gdef\lastpage{\pageref{paper-last-page}}}
\makeatother

\newcommand{\LLMSessions}{10}
\newcommand{\LLMPrompts}{49}
\newcommand{\LLMPromptWords}{1{,}629}
\newcommand{\LLMApiCalls}{1{,}137}
\newcommand{\LLMToolCalls}{1{,}291}

\newcommand{\LLMOutputMTok}{2.2}
\newcommand{\LLMThinkingTokens}{1{,}356{,}149}
\newcommand{\LLMInputMTok}{383.8}
\newcommand{\LLMCacheReadMTok}{375.8}

\newcommand{\LLMCostUSD}{183}

\newcommand{\LLMWorkingHours}{15.0}

\newcommand{\LLMSessionTable}{%
S1 & t+0.0\,h & Algorithm: first version, then sort, overlap, deep stage & 6 & 4.1 & 159 & 50 & 429 & 36 \\
S2 & t+49.8\,h & Finding Pinho et al.'s 2009 code & 3 & 0.1 & 17 & 1 & 8 & 1 \\
S3 & t+50.1\,h & This paper & 14 & 1.8 & 195 & 93 & 380 & 44 \\
S4 & t+61.3\,h & Performance review of mawmetal & 2 & 1.7 & 153 & 61 & 398 & 28 \\
S5 & t+62.5\,h & Benchmark suite; the spruce & 4 & 1.1 & 107 & 23 & 155 & 12 \\
S6 & t+68.0\,h & Larger genomes; download scripts & 1 & 0.1 & 40 & 5 & 39 & 3 \\
S7 & t+68.8\,h & The fern genome: limits & 1 & 0.0 & 6 & 0 & 5 & 0 \\
S8 & t+70.8\,h & Project notes for the agent (/init) & 1 & 0.0 & 9 & 1 & 7 & 1 \\
S9 & t+71.2\,h & List-mode memory; large-genome tuning & 11 & 3.2 & 240 & 92 & 471 & 37 \\
S10 & t+73.6\,h & libsais, libcubwt; tail and deep stage & 6 & 2.8 & 211 & 58 & 305 & 21 \\
\midrule
Total & & & 49 & 15.0 & 1137 & 384 & 2195 & 183 \\}

\newcommand{\BEcoliSpeedup}{11}

\newcommand{\BChrTwentyOneSpeedup}{25}

\newcommand{\BChrOnePinho}{212}

\newcommand{\BChrOneMaw}{306}
\newcommand{\BChrOnePmaw}{261}
\newcommand{\BChrOneEmmaw}{195}
\newcommand{\BChrOneBwt}{56}
\newcommand{\BChrOneMm}{1.1}
\newcommand{\BChrOneMmList}{1.8}
\newcommand{\BChrOneSpeedup}{49}
\newcommand{\BChrOneOverPinho}{186}

\newcommand{\BHumanMm}{29}
\newcommand{\BHumanMmList}{43}

\newcommand{\BHumanMmMem}{10.9}

\newcommand{\BSpruceMm}{221}

\newcommand{\BSpruceMmMem}{17.5}

\newcommand{\BenchMainTable}{%
Human herpesvirus 5 & 0.2 & 0.08 (0.12) & 0.22 & 0.08 & 0.26 & 0.05 & 0.03 (0.03) & 2 \\
\emph{E.~faecalis} V583 & 3.4 & 1.3 (2.0) & 3.3 & 2.9 & 2.7 & 0.57 & 0.06 (0.07) & 10 \\
\emph{E.~coli} K-12 & 4.6 & 1.9 (3.0) & 4.8 & 1.4 & 3.7 & 0.76 & 0.07 (0.08) & 11 \\
\emph{S.~cerevisiae} & 12.2 & 6.3 (9.2) & 13 & 4.1 & 9.7 & 2.1 & 0.13 (0.15) & 16 \\
Human chr21 & 40.1 & 28 (38) & 48 & 42 & 32 & 8.0 & 0.33 (0.43) & 25 \\
Human chr1 & 230.5 & 212 ([33 GB]) & 306 & 261 & 195 & 56 & 1.1 (1.8) & 49 \\
Human GRCh38.p2 & 3{,}061.5 & [168 GB] & [216 GB] & [$>2^{31}$] & [129 GB disk] & [66 GB] & 29 (43) & only \\
Norway spruce & 12{,}018.8 & [642 GB] & [826 GB] & [$>2^{31}$] & [492 GB disk] & [253 GB] & 221 (420) & only \\}
\newcommand{\BenchMemTable}{%
Human herpesvirus 5 & 0.05 & 0.08 & 0.01 & 0.01 & 0.11 & 0.01 & 0.02 & 0.04 \\
\emph{E.~faecalis} V583 & 0.21 & 0.51 & 0.11 & 0.10 & 0.95 & 0.07 & 0.10 & 0.23 \\
\emph{E.~coli} K-12 & 0.28 & 0.69 & 0.15 & 0.13 & 1.33 & 0.10 & 0.13 & 0.25 \\
\emph{S.~cerevisiae} & 0.65 & 1.73 & 0.39 & 0.34 & 2.95 & 0.26 & 0.33 & 0.35 \\
Human chr21 & 2.38 & 5.86 & 1.28 & 1.11 & 5.45 & 0.84 & 0.65 & 0.45 \\
Human chr1 & 12.49 & [33 GB] & 7.44 & 6.35 & 11.15 & 4.79 & 1.35 & 1.01 \\
Human GRCh38.p2 & [168 GB] & [420 GB] & [216 GB] & [$>2^{31}$] & [129 GB disk] & [66 GB] & 10.89 & 9.00 \\
Norway spruce & [642 GB] & [1606 GB] & [826 GB] & [$>2^{31}$] & [492 GB disk] & [253 GB] & 17.55 & 19.87 \\}

\newcommand{\BenchMachineTable}{%
chr21 & Pinho et al. & 119 & 105 & 28 & 26 & 4.3 \\
 & \tool{MAW} & 123 & 73 & 48 & 43 & 2.6 \\
 & \tool{pMAW} & 133 & 75 & 42 & 37 & 3.1 \\
 & \tool{emMAW} & 128 & 57 & 32 & 34 & 4.1 \\
 & \tool{bwtman} & 28 & 14 & 8.0 & 8.0 & 3.5 \\
 & \mm & -- & -- & 0.33 & 0.51 & -- \\
\midrule
chr1 & Pinho et al. & 1{,}456 & 851 & 212 & 180 & 6.9 \\
 & \tool{MAW} & 575 & 484 & 306 & 276 & 1.9 \\
 & \tool{pMAW} & 565 & 468 & 261 & 229 & 2.2 \\
 & \tool{emMAW} & 645 & 339 & 195 & 208 & 3.3 \\
 & \tool{bwtman} & 131 & 86 & 56 & 54 & 2.4 \\
 & \mm & -- & -- & 1.1 & 1.9 & -- \\}
\newcommand{\BChrOneTotalGain}{1{,}277}

\begin{document}

\title{Minimal Absent Words Ten Years Later: Old Computers, New Algorithms, and a Frontier Large Language Model}

\author{Nicolas Innocenti\affil{1}{iProov Sweden AB, Sweden}\affil{2}{Tr\"alhavet Technologies AB, Sweden}}

\contributor{}
\maketitle

\begin{article}

\begin{abstract}
  We designed a new algorithm for computing the minimal absent words (MAWs) of genomes with a large
  language model, Claude, starting from the literature on the problem, and the result outperforms every
  published program we could obtain and run by one to two orders of magnitude. Claude designed the algorithm, implemented it
  for the unified-memory GPU of Apple Silicon computers, optimized and validated it, benchmarked it
  against all earlier programs we could run and wrote this paper, from a short web chat and \LLMPrompts{} prompts totalling \LLMPromptWords{} words and 4 to 8 hours of the author's time, on a
  \$20 monthly subscription. The length distribution of the MAWs of the 12-gigabase Norway spruce
  genome and its reverse complement, which took several days on a 1-terabyte supercomputer in our 2015
  study, now takes under four minutes on a laptop with 32~GB of memory, and the list of all of them seven. No other program we tested can do this on that laptop: in counts mode the in-memory ones would need 253 to 826~GB (1.6~TB for the 2009 program to list the words) or cannot index it, and the external-memory one about 500~GB of free disk. On the human genome the new program reproduces
  our 2015 distribution exactly, all 8,377,095,148 MAWs, in 29 seconds. On the largest genome the
  earlier programs can still process, it is 49 times faster than the fastest of them and 186 times
  faster than the software we used in 2015. We review the algorithms published since 2009, describe the
  new one and what it owes to them, and benchmark every program we could run on four laptops from 2012
  to 2023. Running the same programs on a 2012 laptop and a 2021 one shows that ten years of
  hardware bought a factor of 2 to 7, and asymptotically better algorithms little more. What made the
  difference was fitting the algorithm to the architecture, optimization work that used to be too
  time-consuming to attempt by hand and that a language model now does in hours, for the price of a
  subscription.
\end{abstract}


\abbreviations{
  GPU, graphics processing unit;
  LCP, longest common prefix;
  LLM, large language model;
  MAW, minimal absent word;
  SA, suffix array
}

\dropcap{A} word is absent from a genome if it occurs nowhere in it, and it is a minimal absent word
(MAW) if, in addition, the two words obtained by deleting its first or its last letter both
occur~\cite{Crochemore1998,Pinho2009}. MAWs are a compact basis for the set of all absent words, and
they have been used for fragment assembly~\cite{Mignosi2001,Fici2006}, phylogeny and alignment-free
sequence comparison~\cite{Chairungsee2012,Crochemore2016,Charalampopoulos2018,Saha2026}, the design of
molecular tags that must not match a genome~\cite{Innocenti2015a}, and the search for short sequences
that a pathogen carries and its host lacks~\cite{Silva2015,Pratas2020}. Nullomers, absent words without
the minimality condition, have a literature of their own~\cite{Hampikian2007,Koulouras2021,GeorgakopoulosSoares2021}.

In 2015 we studied the length distribution of the MAWs of thousands of genomes~\cite{Aurell2016}. Its
bulk is what sampling words from a random text produces, while a tail of long MAWs, more than a megabase
long in human, comes from repeated regions shaped by evolution. All the MAWs of that study were computed
with the software of Pinho et al.~\cite{Pinho2009}, with the reverse-complement strand and a length bound of
5,000,000, far above the longest MAW of any genome here. Viruses and bacteria ran on desktop computers and on the author's laptop, a 2012 MacBook Pro.
The human, mouse and Norway spruce~\cite{Nystedt2013} genomes needed Ellen, a 1-terabyte computer at the
PDC Center for High Performance Computing in Stockholm, and the spruce took several days.

This paper is two things at once. It is a new result: an algorithm, \mm{}, that computes the same MAWs
on a laptop, for genomes up to the spruce. And it is a case study in using a large language model (LLM)
to do computational science: the algorithm, its implementation, its optimization, its validation
against every earlier program and the benchmarks were produced by Claude, through the agent Claude
Code~\cite{ClaudeCode}, while the author asked questions, set goals and supplied old code and data.
Even this paper was written by Claude: its text, tables and figures, and the scripts that compute
every number in it from the benchmark results. The case study is not a footnote to the result; the result exists because of it. The author
would not have started this project without an LLM, and the full record of how it was done is part of
the paper (the section \emph{How the work was done} and the Appendix).

The main findings are these. On a 2021 laptop with 32~GB of memory, \mm{} computes the MAWs of the
spruce and its reverse complement, 24 billion bases, in \BSpruceMm~s; no other program we tested fits in its memory and free disk. It
reproduces, length by length, the 8.4 billion human MAWs we computed on Ellen in 2015, in \BHumanMm~s.
On human chromosome~1, the largest genome every program can process, it is \BChrOneSpeedup{} times
faster than the fastest earlier program and \BChrOneOverPinho{} times faster than the 2015 software,
both on the same laptop. Moving the same programs from the 2012 laptop to the 2021 one buys a factor of
2 to 7. And the theoretically better algorithms of the last decade do not explain the gain:
\mm{} is not asymptotically better than its predecessors, and the 2009 software, quadratic in the worst
case, is faster on today's laptop than its linear-time successor.

The lesson we draw goes beyond MAWs. Fitting an algorithm to the machine it runs on (its memory system,
its GPU, its cores) beat ten years of hardware and of algorithmic progress together. We would expect such work to
take a specialist months, and was rarely worth it for a one-off scientific computation. With an LLM it took five days, 4 to 8 hours of the author's spare time, and \$20.

\section{Minimal absent words}

Let $y$ be a string of length $n$ over the alphabet $\Sig=\{A,C,G,T\}$, $\sigma=|\Sig|$. A word is a
factor of $y$ if it occurs in $y$, and absent from $y$ otherwise. An absent word $aWb$, with
$a,b\in\Sig$ and $W\in\Sig^*$, is minimal if both $aW$ and $Wb$ are factors of $y$.

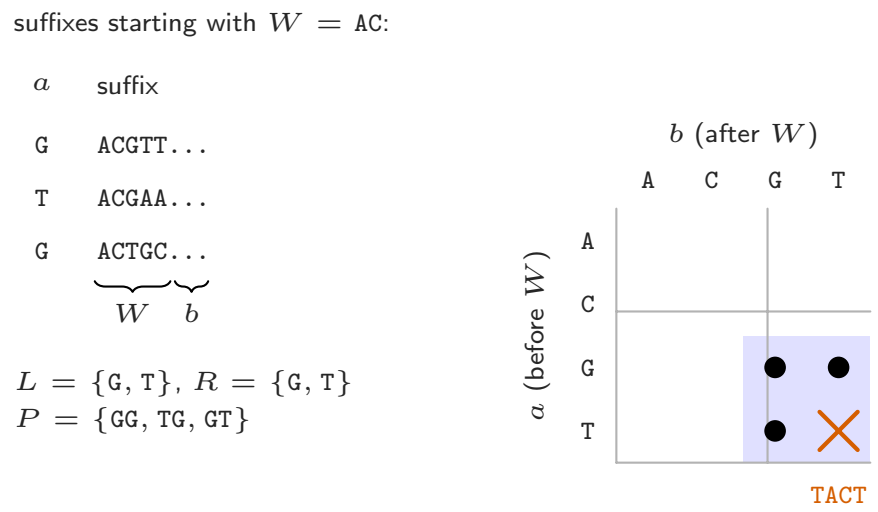
\begin{figure}[t]
  \centering
  \begin{tikzpicture}[font=\sffamily\scriptsize]
    % sorted suffixes sharing W = AC
    \node[anchor=west] at (-0.2,2.55) {suffixes starting with $W=\mathtt{AC}$:};
    \node at (0.1,2.15) {$a$}; \node[anchor=west] at (0.35,2.15) {suffix};
    \foreach \l/\s/\y in {G/ACGTT\dots/1.75, T/ACGAA\dots/1.40, G/ACTGC\dots/1.05} {
      \node at (0.1,\y) {\texttt{\l}}; \node[anchor=west] at (0.35,\y) {\texttt{\s}};
    }
    \draw[decorate, decoration={brace, mirror, amplitude=2pt}] (0.45,0.85) -- (0.95,0.85) node[midway, below=1pt] {$W$};
    \draw[decorate, decoration={brace, mirror, amplitude=2pt}] (0.98,0.85) -- (1.2,0.85) node[midway, below=1pt] {$b$};
    \node[anchor=west, align=left] at (-0.2,0.05) {$L=\{\mathtt{G},\mathtt{T}\}$, $R=\{\mathtt{G},\mathtt{T}\}$\\
      $P=\{\mathtt{GG},\mathtt{TG},\mathtt{GT}\}$};
    % the pair matrix
    \begin{scope}[xshift=3.9cm, yshift=-0.35cm, x=0.42cm, y=0.42cm]
      \foreach \b [count=\j from 0] in {A,C,G,T} { \node at (\j+0.5,4.45) {\texttt{\b}}; }
      \foreach \a [count=\i from 0] in {A,C,G,T} { \node at (-0.45,3.5-\i) {\texttt{\a}}; }
      \node at (2,5.15) {$b$ (after $W$)};
      \node[rotate=90] at (-1.25,2) {$a$ (before $W$)};
      \fill[blue!12] (2,0) rectangle (4,2);
      \draw[black!30] (0,0) grid (4,4);
      \fill (2.5,1.5) circle (0.17); \fill (2.5,0.5) circle (0.17); \fill (3.5,1.5) circle (0.17);
      \draw[okVermillion, thick] (3.2,0.2) -- (3.8,0.8) (3.2,0.8) -- (3.8,0.2);
      \node[okVermillion, anchor=north] at (3.5,-0.1) {$\mathtt{TACT}$};
    \end{scope}
  \end{tikzpicture}
  \caption{The MAWs of one node. The suffixes that start with $W$ give the bases before ($L$) and after
    ($R$) its occurrences, and the pairs $P$ that occur together (dots). Every pair in $L\times R$
    (shaded) that does not occur is a MAW $aWb$: here only $\mathtt{TACT}$, since $\mathtt{TAC}$ and
    $\mathtt{ACT}$ occur but $\mathtt{TACT}$ does not. With a fifth symbol $\none$ for ``none'' (the
    start or end of a segment), $L$, $R$ and $P$ fit in 5, 5 and 25 bits.
    \label{fig:node}}
\end{figure}

Two facts underlie every algorithm below. First, the infix $W$ of a MAW occurs at least twice in $y$:
once inside an occurrence of $Wb$ and once inside an occurrence of $aW$, and the two occurrences differ
because $aWb$ is absent. Second, $W$ is a maximal repeat: it is preceded by $a$ and, somewhere, by
another letter or by the start of $y$, and followed by $b$ and, somewhere, by another letter or by the
end of $y$. In the suffix tree of $y$, $W$ is therefore an internal node, and the MAWs with infix $W$
are the pairs $(a,b)$ of a left and a right extension of $W$ that never occur together
(Fig.~\ref{fig:node}). The number of MAWs is $O(\sigma n)$~\cite{Crochemore1998,Mignosi2002}, and this
bound is tight~\cite{Charalampopoulos2018}.

A genome is not one string. It has several chromosomes or contigs, two strands, and letters other than
A, C, G and T, mostly runs of N. As in 2015, we take the forward sequence of every record and its
reverse complement, and cut them at every other letter into \emph{segments}; a word occurs if it occurs
inside a segment. The MAWs of a genome are those of its set of segments, taken together.

\section{Ten years of algorithms}

Table~\ref{tab:algorithms} lists the algorithms for computing MAWs published since the
1990s; we describe them roughly in order.

\begin{figure*}[t]
  \tabcaption{tab:algorithms}{Algorithms for computing all minimal absent words of a string of length $n$ over an
    alphabet of size $\sigma$. Space in machine words unless stated. For \mm{}, $D=1024$ is the depth of
    its deep stage and $w=32$ the number of bases in a 64-bit word. Code: ``run'', benchmarked here;
    ``paper'', the paper reports an implementation that is not public or does not compute MAWs; --, none
    found.}
  \centering\footnotesize
  \begin{tabular}{@{}lp{0.25\textwidth}p{0.27\textwidth}p{0.28\textwidth}l@{}}
    \toprule
    Year & Reference & Approach & Time; working space & Code \\
    \midrule
    1998 & Crochemore, Mignosi, Restivo~\cite{Crochemore1998} & suffix automaton (DAWG) & $O(\sigma n)$; $O(n)$ & -- \\
    2008 & Herold, Kurtz, Giegerich~\cite{Herold2008} & enhanced suffix array (shortest absent words only) & $O(n)$; $O(n)$ & paper \\
    2009 & Pinho et al.~\cite{Pinho2009} & suffix array, LCP intervals & $O(n^2)$ worst case; $O(n)$ & run \\
    2013--20 & Belazzougui, Cunial et al.\ (\tool{bwtman})~\cite{Belazzougui2013,Belazzougui2017,Belazzougui2020} & bidirectional BWT, maximal repeats & $O(n)$; $O(n\log\sigma)$ bits & run \\
    2014 & Barton et al.\ (\tool{MAW})~\cite{Barton2014} & suffix array, LCP intervals & $O(n)$; $O(n)$ & run \\
    2016 & Barton et al.\ (\tool{pMAW})~\cite{Barton2015} & \tool{MAW} in parallel (OpenMP) & $O(n)$ work; $O(n)$ & run \\
    2016 & Fujishige et al.~\cite{Fujishige2016,Fujishige2023} & DAWG for integer alphabets & $O(n+|\mathrm{MAW}|)$; $O(n)$ & -- \\
    2017 & H\'eliou, Pissis, Puglisi (\tool{emMAW})~\cite{Heliou2017} & suffix and LCP arrays in external memory & $O(n)$; RAM chosen by the user, $O(n)$ disk & run \\
    2020 & Crochemore et al.~\cite{Crochemore2020}, Mieno et al.~\cite{Mieno2020} & sliding window, online & see references & -- \\
    2021 & Ayad et al.~\cite{Ayad2021} & $k$ documents, output-sensitive space & $O(kn+\sum|M^\ell|)$; $O(\mathrm{MaxIn}+\mathrm{MaxOut})$ & paper \\
    2021 & Nishimoto, Tabei (r-enum)~\cite{Nishimoto2021} & run-length BWT ($r$ runs) & $O(n\log\log(n/r))$; $O(r\log n)$ bits & paper \\
    2022 & Akagi et al.~\cite{Akagi2022} & run-length encoding ($m$ runs) & $O(|\mathrm{MAW}|)$ reporting; $O(m)$ & -- \\
    2023 & Okabe et al.~\cite{Okabe2023} & generalized MAWs of several strings & $O(n+|\mathrm{MAW}|)$; $O(n)$ & -- \\
    2024 & Inenaga et al.~\cite{Inenaga2024} & CDAWG ($e$ edges) & $O(|\mathrm{MAW}|)$; $O(e)$ & -- \\
    2026 & this work (\mm) & GPU radix sort of suffix keys; suffix array of long repeats only & $O(nD/w)$ worst case, plus the output in list mode; $O(n)$; about $0.5n$ bytes on genomes, up to about $9n$; DNA only & run \\
    \bottomrule
  \end{tabular}
\end{figure*}

\subsection{The state of the art in 2009}
Crochemore, Mignosi and Restivo~\cite{Crochemore1998} gave the first linear-time algorithm, built on the
suffix automaton (the directed acyclic word graph, DAWG). Its memory, tens of bytes per base, kept it
away from genomes. Herold, Kurtz and Giegerich~\cite{Herold2008} computed the shortest absent words
with an enhanced suffix array, and Wu, Jiang and Su~\cite{Wu2010} gave another algorithm for that case.
Pinho et al.~\cite{Pinho2009} were the first to compute all MAWs of genomes. They build a suffix array
with the Quinlan--Dorward suffix sorter distributed by McIlroy~\cite{McIlroy}, an LCP array with the
algorithm of Kasai et al.~\cite{Kasai2001}, and find the MAWs while traversing the LCP intervals with a
stack. Barton et al.\ point out that this takes quadratic time in the worst case~\cite{Barton2014}, but
it was fast in practice, and Garcia and Pinho used it on human assemblies~\cite{Garcia2011a,Garcia2011b}.
It is the software of our 2015 study.

\subsection{Linear time, in parallel and on disk}
Barton, H\'eliou, Mouchard and Pissis~\cite{Barton2014} gave an $O(n)$-time, $O(n)$-space algorithm in
2014. A bottom-up traversal of the LCP intervals visits every internal node of the suffix tree; two bit
vectors per node record which letters precede and follow its occurrences, and the MAWs are read off from
them. Their \tool{MAW} builds the suffix array with libdivsufsort~\cite{divsufsort}, outperformed Pinho et
al.'s implementation on their data, and has remained the tool applied work uses~\cite{Koulouras2021}. The
same group parallelised it with OpenMP (\tool{pMAW})~\cite{Barton2015}. H\'eliou, Pissis and
Puglisi's \tool{emMAW}~\cite{Heliou2017} builds the suffix and LCP arrays in external memory; it
processes a human genome in under 3 hours with 1~GB of RAM, and is less than twice as slow as the
in-memory tools when memory is plentiful.

\subsection{Small and compressed space}
Since 2013 the theory has gone toward smaller space. Belazzougui, Cunial, K\"arkk\"ainen and
M\"akinen~\cite{Belazzougui2013,Belazzougui2020} enumerate the internal nodes of the suffix tree, and with
them the maximal repeats and the MAWs, in $O(n)$ time and $O(n\log\sigma)$ bits, from a bidirectional
Burrows--Wheeler transform (BWT); Belazzougui and Cunial~\cite{Belazzougui2017} used it for string
kernels, and Cunial's prototype, \tool{bwtman}~\cite{Cunial}, computes MAWs on all cores. Fujishige et
al.~\cite{Fujishige2016,Fujishige2023} extended the DAWG approach to integer alphabets. Ayad et
al.~\cite{Ayad2021} work in space proportional to the largest input and output. For collections of
similar genomes, r-enum~\cite{Nishimoto2021} works on the run-length encoded BWT, and Akagi et
al.~\cite{Akagi2022} and Inenaga et al.~\cite{Inenaga2024} report MAWs from run-length encodings and
compact DAWGs. MAWs have also been computed for a sliding window~\cite{Crochemore2020,Mieno2020}, for
several strings at once~\cite{Okabe2023}, and as the basis of sequence
comparison~\cite{Crochemore2016,Charalampopoulos2018}.

\subsection{Where things stood in 2026}
Five of these programs could be obtained and run: Pinho et al.'s, \tool{MAW}, \tool{pMAW}, \tool{emMAW}
and \tool{bwtman}. The public r-enum code enumerates maximal repeats and minimal unique substrings, not
MAWs; Ayad et al.'s implementation is available on request; we found no code for the others. Suffix
sorting, which dominates the suffix-array programs, has kept improving: libsais~\cite{libsais}, an
engineered implementation of induced sorting~\cite{Nong2011}, is several times faster than
libdivsufsort and parallel. Suffix arrays have been built on GPUs~\cite{Deo2013,Wang2016,Buren2019},
but we found no MAW program that uses one.

\section{What is special about Apple Silicon}

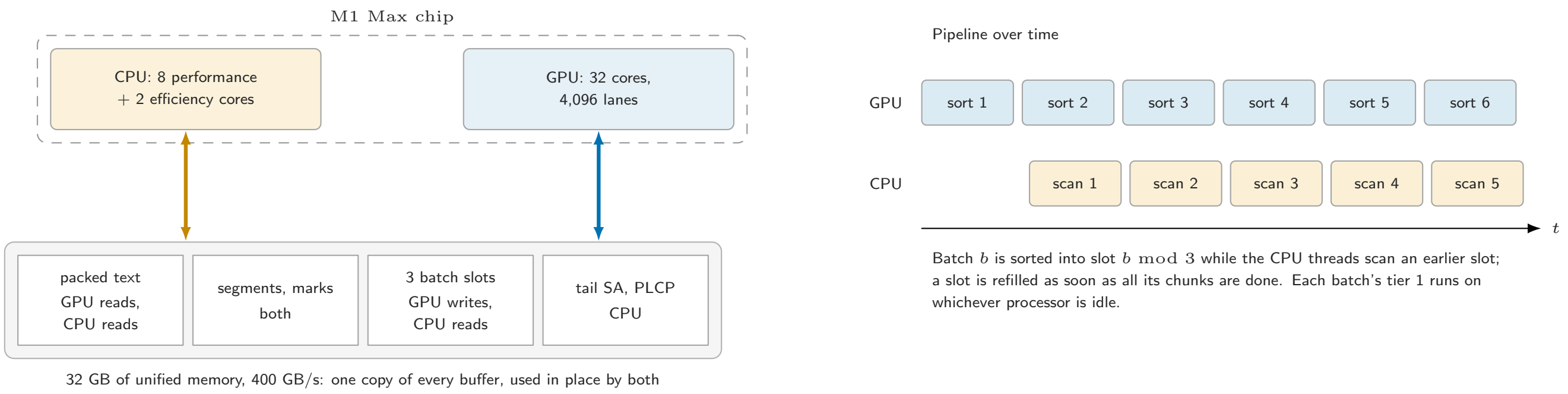
\begin{figure*}[t]
  \centering
  \begin{tikzpicture}[font=\sffamily\scriptsize,
      blk/.style={draw=black!35, rounded corners=2pt, align=center, inner sep=3pt, minimum height=9mm},
      buf/.style={draw=black!35, fill=white, align=center, inner sep=2pt, text width=17mm, minimum height=10mm},
      arr/.style={{Latex[length=1.6mm]}-{Latex[length=1.6mm]}, very thick}]
    \node[blk, fill=okOrange!14, text width=28mm] (cpu) at (0,0) {CPU: 8 performance\\+ 2 efficiency cores};
    \node[blk, fill=okBlue!10, text width=28mm] (gpu) at (4.6,0) {GPU: 32 cores,\\4{,}096 lanes};
    \node[draw=black!40, dashed, rounded corners=3pt, fit=(cpu)(gpu), inner sep=4pt, label={[font=\scriptsize]above:M1 Max chip}] (chip) {};
    \node[buf] (text) at (-0.95,-2.35) {packed text\\[1pt]{\tiny GPU reads, CPU reads}};
    \node[buf] (marks) at (1.0,-2.35) {segments, marks\\[1pt]{\tiny both}};
    \node[buf] (slots) at (2.95,-2.35) {3 batch slots\\[1pt]{\tiny GPU writes, CPU reads}};
    \node[buf] (tail) at (4.9,-2.35) {tail SA, PLCP\\[1pt]{\tiny CPU}};
    \node[draw=black!30, rounded corners=3pt, fill=black!4, fit=(text)(marks)(slots)(tail), inner sep=4pt] (mem) {};
    \node[buf] at (text) {packed text\\[1pt]{\tiny GPU reads, CPU reads}};
    \node[buf] at (marks) {segments, marks\\[1pt]{\tiny both}};
    \node[buf] at (slots) {3 batch slots\\[1pt]{\tiny GPU writes, CPU reads}};
    \node[buf] at (tail) {tail SA, PLCP\\[1pt]{\tiny CPU}};
    \node[below=1pt of mem] {32~GB of unified memory, 400~GB/s: one copy of every buffer, used in place by both};
    \draw[arr, okOrange!85!black] (cpu.south) -- (cpu.south |- mem.north);
    \draw[arr, okBlue] (gpu.south) -- (gpu.south |- mem.north);
    % timeline
    \begin{scope}[xshift=82mm, yshift=-1.6cm]
      \node[anchor=west] at (0,2.2) {Pipeline over time};
      \node[anchor=east] at (-0.1,1.45) {GPU};
      \node[anchor=east] at (-0.1,0.55) {CPU};
      \foreach \i in {1,...,6} \draw[draw=black!35, rounded corners=1.5pt, fill=okBlue!14] (\i*1.12-1.12,1.2) rectangle ++(1.02,0.5) node[midway] {sort \i};
      \foreach \i in {1,...,5} \draw[draw=black!35, rounded corners=1.5pt, fill=okOrange!16] (\i*1.12+0.08,0.3) rectangle ++(1.02,0.5) node[midway] {scan \i};
      \draw[-{Latex[length=1.5mm]}] (0,0.05) -- (6.9,0.05) node[right] {$t$};
      \node[anchor=north west, align=left, text width=66mm] at (0,-0.1) {Batch $b$ is sorted into slot
        $b \bmod 3$ while the CPU threads scan an earlier slot; a slot is refilled as soon as all its
        chunks are done. Each batch's tier~1 runs on whichever processor is idle.};
    \end{scope}
  \end{tikzpicture}
  \caption{Apple Silicon as \mm{} uses it. Left: the CPU and the GPU share one memory, so the GPU sorts
    suffix keys into buffers that the CPU scans in place, and both read the same packed genome. Right:
    three batch slots keep both busy.\label{fig:arch}}
\end{figure*}

\mm{} is written for one family of computers, Apple's laptops and desktops since 2020, and it is fast
because it uses what is unusual about them (Fig.~\ref{fig:arch}). On the 2021 MacBook Pro used here, an
M1 Max chip holds 10 CPU cores (8 performance and 2 efficiency cores), a GPU with 32 cores and 4,096
arithmetic units, and 32~GB of memory that both share, at 400~GB/s~\cite{AppleM1Max}.

\emph{Unified memory} is the important part. A GPU in a desktop computer has its own memory: data must be
copied to it over a bus that is many times slower than the GPU's own memory, and back, so GPU programs
are designed to cross that bus as little as possible, and a genome-sized index rarely fits. On Apple
Silicon the CPU and the GPU read and write the same memory pages. A Metal buffer can wrap an ordinary
page-aligned allocation without copying it, so the GPU sorts data the CPU has just written, and the CPU
reads the GPU's output in place. Work can move between the two at the granularity of a
batch of a few million suffixes, without a copy in either direction.

The two processors are also unlike each other in useful ways. The GPU runs threads in SIMD groups of
32 that can vote (\texttt{simd\_ballot}) and share 32~KB of threadgroup memory, which suits radix sorting
and scans of fixed-width keys; unlike NVIDIA's GPUs, it does not guarantee that a threadgroup waiting
for another one makes progress, which rules out the single-pass ``decoupled look-back'' scans of the
fastest GPU sorts~\cite{Adinets2022}. The CPU cores are few, fast, and have large caches, which suits the
irregular, branchy work that GPUs do badly; they are of two kinds, so a parallel step that waits for its
slowest thread should not run on the efficiency cores. Finally, there are limits: a single Metal buffer
can hold at most about 20~GB on this machine, and the GPU's recommended working set is about two thirds
of the memory.

These features shaped the algorithm. The bulk of the work, sorting short keys of every position, goes to
the GPU, where it is a few uniform passes over memory; the irregular parts (scanning the sorted keys,
refining groups of repeats) go to the CPU cores; and neither ever copies data to the other. The batches
are sized to the GPU's working set, three batch slots let the GPU sort while the CPU scans, the
tier-1 scan runs on whichever processor is idle, the input is packed with the CPU's NEON vector
instructions, the suffix-array library runs on the performance cores only, and data structures are laid
out for the caches and the 16~KB memory pages. None of this changes what is computed; all of it changes
how fast.

\section{An algorithm for a laptop}

\subsection{The idea: split by the length of the infix}
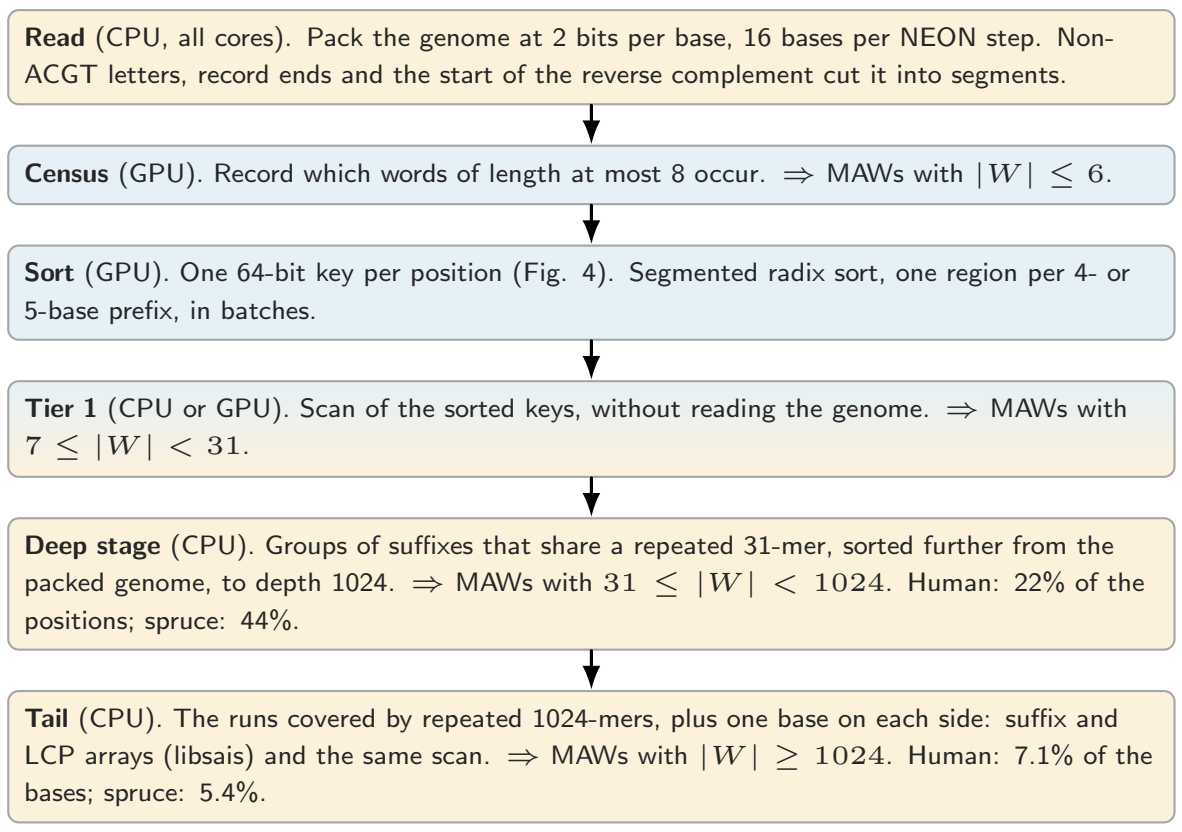
\begin{figure}[t]
  \centering
  \begin{tikzpicture}[
      font=\sffamily\scriptsize,
      box/.style={draw=black!35, rounded corners=2pt, align=left, text width=0.86\columnwidth, inner sep=3pt},
      gpu/.style={box, fill=okBlue!10},
      cpu/.style={box, fill=okOrange!14},
      both/.style={box, top color=okBlue!10, bottom color=okOrange!14},
      arr/.style={-{Latex[length=1.6mm]}, semithick},
      node distance=2.6mm]
    \node[cpu] (pack) {\textbf{Read} (CPU, all cores). Pack the genome at 2 bits per base, 16 bases per
      NEON step. Non-ACGT letters, record ends and the start of the reverse complement cut it into
      segments.};
    \node[gpu, below=of pack] (census) {\textbf{Census} (GPU). Record which words of length at most 8 occur.
      $\Rightarrow$ MAWs with $|W|\le 6$.};
    \node[gpu, below=of census] (sort) {\textbf{Sort} (GPU). One 64-bit key per position (Fig.~\ref{fig:key}).
      Segmented radix sort, one region per 4- or 5-base prefix, in batches.};
    \node[both, below=of sort] (scan) {\textbf{Tier 1} (CPU or GPU). Scan of the sorted keys, without reading
      the genome. $\Rightarrow$ MAWs with $7\le|W|<31$.};
    \node[cpu, below=of scan] (deep) {\textbf{Deep stage} (CPU). Groups of suffixes that share a repeated
      31-mer, sorted further from the packed genome, to depth 1024. $\Rightarrow$ MAWs with $31\le|W|<1024$.
      Human: 22\% of the positions; spruce: 44\%.};
    \node[cpu, below=of deep] (res) {\textbf{Tail} (CPU). The runs covered by repeated 1024-mers, plus one
      base on each side: suffix and LCP arrays (libsais) and the same scan. $\Rightarrow$ MAWs with
      $|W|\ge 1024$. Human: 7.1\% of the bases; spruce: 5.4\%.};
    \foreach \a/\b in {pack/census, census/sort, sort/scan, scan/deep, deep/res} \draw[arr] (\a) -- (\b);
  \end{tikzpicture}
  \caption{The tiers of \mm{}. Blue: GPU; orange: CPU. Each tier handles the MAWs $aWb$ whose infix length
    lies in its range, and only positions that start repeats of the previous tier's depth reach the next
    one. Shares: GRCh38.p2 and the spruce, both strands.\label{fig:pipeline}}
\end{figure}

\mm{} starts from the two facts above: the infix $W$ of every MAW $aWb$ is a repeat, and the MAWs of a
genome are overwhelmingly short~\cite{Aurell2016}. Short MAWs only require comparing short prefixes of
the suffixes, which fit in a machine word; long MAWs require long repeats, which cover a small part of a
genome. The algorithm therefore handles the MAWs in tiers by $|W|$ (Fig.~\ref{fig:pipeline}): each tier
is more expensive per position than the one before, and receives only the positions that the one before
could not settle.

\subsection{Text and keys}
\begin{figure}[t]
  \centering
  \begin{tikzpicture}[font=\sffamily\scriptsize, x=1.15mm, y=1mm]
    \draw[draw=black!35, fill=okBlue!10] (0,0) rectangle (56,6) node[midway] {bases $m+1\ldots m+28$ of the suffix, 2 bits each};
    \draw[draw=black!35, fill=okOrange!14] (56,0) rectangle (61,6) node[midway] {len};
    \draw[draw=black!35, fill=okGreen!14] (61,0) rectangle (64,6) node[midway] {$a$};
    \foreach \x/\l in {0/63, 56/7, 61/2, 64/0} \node[below] at (\x,0) {\l};
    \node[anchor=west, align=left] at (0,-8) {len: valid bases (5 bits); $a$: the preceding base, or ``none'' (3 bits).\\
      Only the bases up to depth 31 are sorted, $2(31-m)$ bits in six radix passes:\\
      six of 9 bits for $m=4$ (54 bits); five of 9 and one of 7 for $m=5$ (52 bits).};
  \end{tikzpicture}
  \caption{The 64-bit key of a position. The first $m$ bases are implicit in the key's region.\label{fig:key}}
\end{figure}
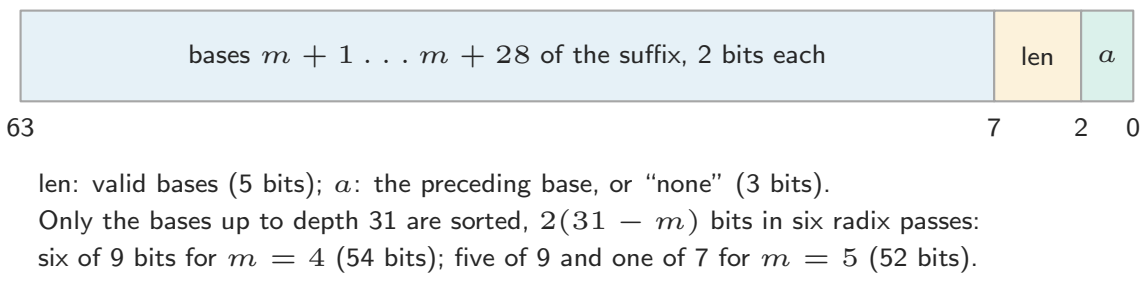

The genome is read once, by all cores, and packed at two bits per base, sixteen bases per 32-bit word,
with a bitmap marking where segments start. The reverse complement is built from the packed bases.
Every position $p$ then gets a 64-bit key (Fig.~\ref{fig:key}): the 28 bases that follow the first $m$
bases of the suffix at $p$, how many of them are valid before the segment ends, and the base that
precedes $p$. Positions are grouped into $4^m$ \emph{regions} by their first $m$ bases, with $m=4$, or
$m=5$ on texts of a gigabase or more, where smaller regions keep the sort within the GPU's caches.

Sorting the keys of a region on their top $2(31-m)$ bits orders its suffixes by their first 31 bases,
which is their suffix-array order to that depth and all that tier~1 needs, since suffixes that share 31 bases are left to the deep stage. Two precautions
are needed. A suffix that ends inside the key is padded with zeros, which
would compare like a run of A; so the valid length is kept, and the few suffixes with fewer valid bases
than the key holds are first ordered by that length, so that the stable sort places a suffix that ends
early before its extensions, as a terminating sentinel would. And the LCP of two adjacent keys is the
number of leading zero bit pairs of their exclusive or, capped by their valid lengths.

\subsection{Short MAWs: a census}
One GPU pass records which words of length at most 8 occur inside segments; with them, every word $aWb$
with $|W|\le6$ is a MAW exactly when it is absent and $aW$ and $Wb$ are present. This settles the very
short MAWs, and it also means that tier~1 never reports a node shallower than 7 bases, so its work can be
cut into chunks wherever adjacent suffixes share fewer than 7 bases, every few suffixes, and spread
evenly over all cores.

\subsection{Sorting on the GPU}
The keys of a batch of regions are built on the GPU from the packed text and sorted by a segmented
least-significant-digit radix sort~\cite{Satish2009,Merrill2011}. Tiles of 2048 keys never straddle a
region. Each pass counts the digits of every tile, finds the destination of every (region, digit, tile)
triple with one exclusive scan~\cite{Blelloch1990,Sengupta2007}, ranks each tile's keys stably with SIMD-group
ballots, reorders them in threadgroup memory, and writes each digit's run with consecutive threads on
consecutive addresses. The positions travel with the keys, as 32-bit integers plus, above $2^{32}$
bases, a separate high byte: five bytes per position, for inputs of up to 68.7~Gbp.

\subsection{Tier 1: scanning the sorted keys}
The CPU traverses the LCP intervals of each sorted chunk bottom-up~\cite{Kasai2001,Abouelhoda2004,Barton2014},
but from the keys alone: since a key carries the base before its position and the bases after it, the
scan never reads the genome. For an interval $v$ of string depth $\ell=|W|$, each child $c$, a
sub-interval or a single suffix, contributes the set $L_c$ of bases before its suffixes and the base
$\beta_c$ after $W$ in them ($\none$ for a suffix that ends at depth $\ell$). Then
\begin{equation}
  L_v=\bigcup_c L_c,\qquad R_v=\{\beta_c\},\qquad P_v=\bigcup_c L_c\times\{\beta_c\},
\end{equation}
and the MAWs with infix $W$ are
\begin{equation}
  \{\,aWb \;:\; a\in L_v\cap\Sig,\ b\in R_v\cap\Sig,\ (a,b)\notin P_v\,\}
  \label{eq:node}
\end{equation}
(Fig.~\ref{fig:node}). With five symbols, a node costs a few word operations per child.

In counts mode the GPU can do the same work differently. For each depth $d$ from 7 to 30, the suffixes
that share their first $d$ bases form a run of the sorted keys; the run's matrix of (preceding base,
base at depth $d$) pairs is the OR of one bit per suffix, and its MAWs follow from it as in
Eq.~\ref{eq:node}. One segmented OR scan finds the runs of all 24 depths at once. Each batch's tier~1
runs on whichever processor is less busy when the batch is submitted.

\subsection{The deep stage}
Every occurrence of a $W$ with 31 or more bases starts with the same 31 bases, so all of them lie in one
group of tier~1's sorted output. The deep stage sorts each group further, reading 32 bases at a time from
the packed genome, down to depth $D=1024$, and the same scan gives the MAWs with $31\le|W|<1024$. Groups
average about five members; a group of two (60\% of the groups on the human genome) needs only one
comparison. The reads, random accesses into a genome of several gigabytes, are prefetched well ahead,
inside long exact repeats several windows are compared at once, and a coarse bitmap of segment starts
answers most ``where does this segment end'' questions from the cache. Positions still tied at depth
$D$ start a repeated $D$-mer and are marked. Refining further would be quadratic inside long exact
repeats, where a suffix array is linear.

\subsection{The tail}
The positions covered by marked $D$-mers are merged into runs, each extended by one base on each side
within its segment, and concatenated with separators. libsais~\cite{libsais} builds the generalized suffix
array of this \emph{residue} and its permuted LCP array~\cite{Karkkainen2009}, and the same scan, reading
each LCP through the suffix array, gives the MAWs with $|W|\ge D$: nine bytes per residue base. Before
that, the batch buffers and, for large residues, the packed text are freed.

\subsection{Exactness}
The tiers partition the MAWs by $|W|$. The census and the scans are exact by construction; the deep
stage and the tail rest on the following statement. With $t=31$, its first step is why a group holds
every occurrence of each $W$ the deep stage handles; with $t=D$, it makes the tail exact.

\noindent\textbf{Proposition.} Let $t\ge2$, let $C_t$ be the set of positions covered by occurrences of
$t$-mers that occur at least twice in the genome (an occurrence at $p$ covers $p,\ldots,p+t-1$), and let the residue consist of the maximal runs of
$C_t$, each extended by one base on each side within its segment and followed by a separator, a symbol not in $\Sig$. A word
$aWb$ with $|W|\ge t$ is a MAW of the genome if and only if it is a MAW of the residue.

\noindent\textit{Proof.} Words of the residue that contain no separator are words of the genome. ($\Rightarrow$)
Let $aWb$ be a MAW of the genome. $W$ occurs twice, so every $t$-mer of $W$ occurs twice and every
occurrence of $W$ lies inside a run. The extension keeps the letters before and after, which lie in the
same segment, so $aW$ and $Wb$ occur in the residue; $aWb$ does not, since it does not occur in the
genome. ($\Leftarrow$) Let $aW$ and $Wb$ occur in the residue and $aWb$ not; then $aW$ and $Wb$ occur in
the genome. Suppose $aWb$ occurs in the genome. If $W$ occurs at least twice, this occurrence of $W$ lies
inside a run, whose extension contains $aWb$: a contradiction. If $W$ occurs once, the occurrences of
$aW$ and $Wb$ in the residue both come from that occurrence; maximal runs are at least one position
apart, so two extended runs share at most one position, while $aW$ and $Wb$ share all of $W$, at least two
positions, so they lie in the same extended run, which contains $aWb$: a contradiction again. $\square$

\subsection{Counts mode and list mode}
Two kinds of output are useful, and they cost very differently. In \emph{counts mode}, the default,
\mm{} reports how many MAWs there are of each length, which is what our 2015 study used. Every MAW is still
found, one by one, as a missing pair at its node (Eq.~\ref{eq:node}), but it is only counted: the number
of missing pairs is added to the histogram at length $|W|+2$, and no word is ever spelled out. The
output is a few kilobytes, and tier~1 can run on the GPU. In \emph{list mode} (\texttt{--list}), every MAW
is written out as text, one word per line, as \tool{MAW} does: the CPU spells $a$, $W$ and $b$ from the
key or the packed genome, chunks are written in a fixed order, so finished chunks wait for their turn,
and the output is enormous. The list of human chromosome~1 with its reverse complement is about 13~GB of
text, that of the human genome about 170~GB, that of the spruce 498~GB, and a single thread writing to
memory tops out at 6.4~GB/s on this laptop. List mode is therefore CPU- and output-bound. The earlier
programs come in both kinds: Pinho et al.'s and \tool{bwtman} report counts (Pinho et al.'s writes words
when asked), \tool{MAW} and \tool{emMAW} always write words, and \tool{pMAW} writes one file per thread.
The benchmarks report both modes where a program has both, and every speedup quoted compares counts
with counts or lists with lists.

\subsection{What was borrowed and what is new}
\mm{} is mostly made of known parts, and it is worth saying which (Table~\ref{tab:origins}). The facts it
rests on, that MAW infixes are repeats and maximal repeats~\cite{Crochemore1998,Belazzougui2013}, are
classical. Its scan is the LCP-interval traversal of Kasai et al.\ and Abouelhoda et al.~\cite{Kasai2001,Abouelhoda2004}
with the per-node extension sets of Barton et al.~\cite{Barton2014}. Radix sorting of fixed-width keys,
segmented scans and SIMD-group ranking are standard GPU techniques~\cite{Blelloch1990,Sengupta2007,Satish2009,Merrill2011};
GPU suffix-array builders also sort fixed-width prefixes by radix~\cite{Deo2013,Wang2016}. The deep stage is
a bounded refinement of buckets, as in multikey quicksort~\cite{BentleySedgewick1997}, and the switch to
a different method for long repeats is the ``deep--shallow'' strategy of Manzini and
Ferragina~\cite{ManziniFerragina2004}. The tail uses libsais and the permuted LCP array~\cite{libsais,Karkkainen2009}.

What we could not find in the literature, and what Claude proposed, is the combination: splitting the
MAWs by infix length into tiers that move from the GPU to the CPU and from short keys to a suffix array;
restricting the suffix array to the regions covered by repeats of the depth where the cheaper tiers stop,
with the proof that this is exact; a key that carries the preceding base and its valid length, so that
every scan works on keys alone; the segmented-OR formulation of tier~1, which finds the nodes of 24 depths
in one GPU scan; and the engineering for unified memory, where the stages share buffers instead of
copying them and tier~1 moves between processors batch by batch. The tiered design appeared in the first
claude.ai conversation, in reply to the author's objection that the longest MAWs are very long; the deep
stage, in reply to the question of what the spruce would need.

\begin{figure}[t]
  \tabcaption{tab:origins}{Where the parts of \mm{} come from.}
  \centering\footnotesize
  \begin{tabular}{@{}p{0.40\columnwidth}p{0.53\columnwidth}@{}}
    \toprule
    Part & Origin \\
    \midrule
    MAW infixes are maximal repeats & \cite{Crochemore1998,Belazzougui2013} \\
    Node scan, extension sets & \cite{Kasai2001,Abouelhoda2004,Barton2014} \\
    GPU radix sort, scans, ranking & \cite{Blelloch1990,Sengupta2007,Satish2009,Merrill2011} \\
    Sorting fixed-width prefixes & GPU suffix arrays~\cite{Deo2013,Wang2016} \\
    Deep stage, cut at depth $D$ & \cite{BentleySedgewick1997,ManziniFerragina2004} \\
    Tail: SA, permuted LCP & libsais, \cite{libsais,Karkkainen2009} \\
    Tiers by $|W|$; residue and its proof & new (Claude) \\
    Keys with preceding base and length & new (Claude) \\
    Segmented-OR tier 1 on the GPU & new (Claude) \\
    Unified-memory pipeline, load balancing & new (Claude) \\
    \bottomrule
  \end{tabular}
\end{figure}

\subsection{Implementation and validation}
\begin{figure}[t]
  \centering
  \input{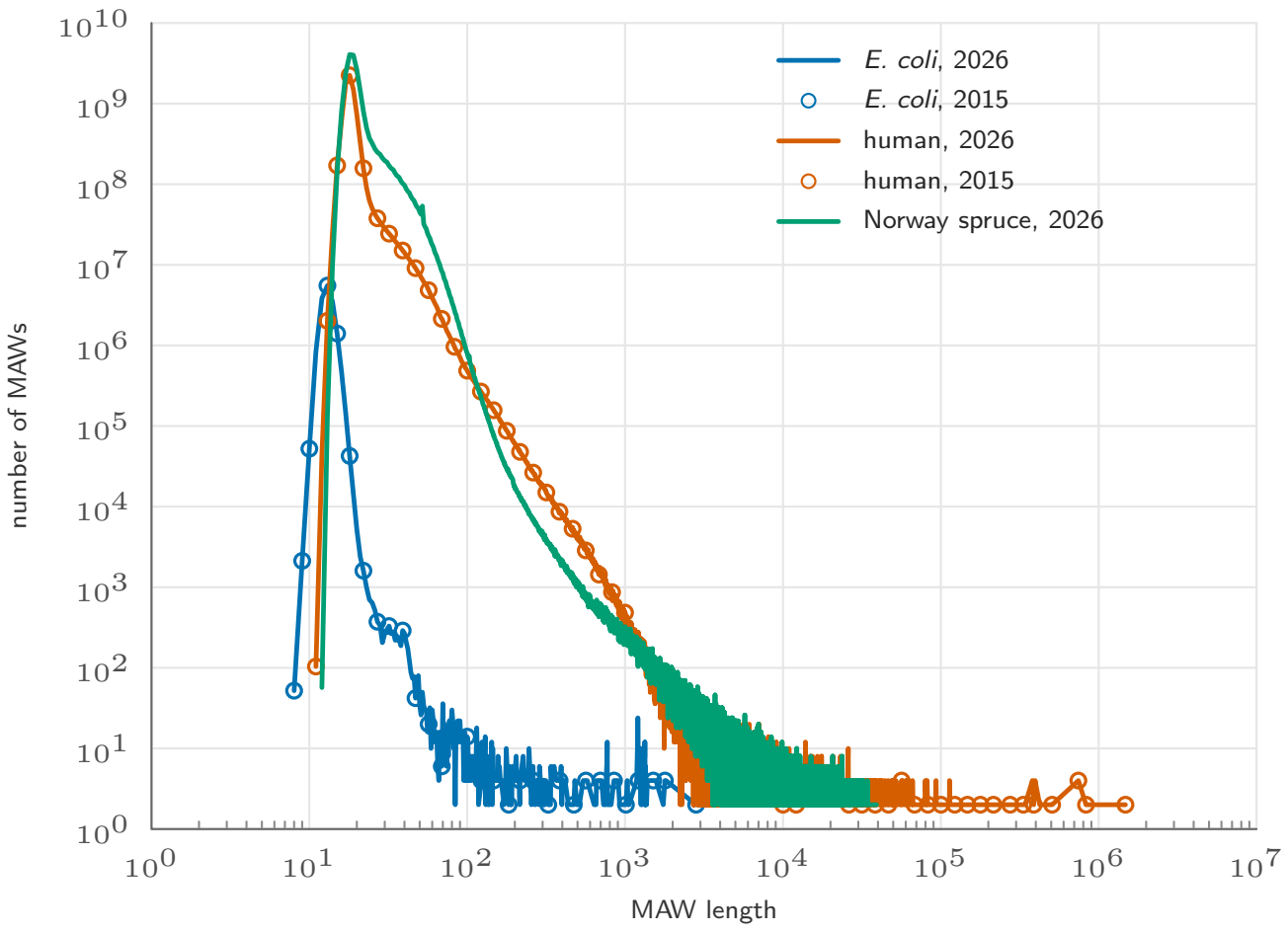}
  \caption{MAW length distributions: our 2015 files (circles, thinned) and today's \mm{} runs (lines),
    both strands. The 2015 and 2026 counts are identical at every length, for \emph{E.~coli} and for the
    human genome (8.4 billion MAWs, 7,822 lengths). The spruce was not available from 2015: 21.2 billion
    MAWs, the longest 39,131 bases.\label{fig:dist}}
\end{figure}

\mm{} is written in C++17, Objective-C++ for the Metal host code and the Metal Shading Language for the
GPU kernels, and released under the MIT license. It accepts the options of \tool{MAW}, and
\texttt{--joint} computes the MAWs of all records together, as Pinho et al.'s program does. Its output
agrees with every earlier program except \tool{pMAW}, whose errors we checked by brute force. A randomized differential test compares it with a brute-force
reference on inputs that mix planted and mutated repeats, homopolymers, N runs, lowercase letters and
many tiny records, and randomizes every tier boundary, batch size, chunk size and output path so that
small inputs exercise every stage; the last full runs passed $3\times300$ cases. On the four smallest
genomes of Table~\ref{tab:genomes}, Pinho et al.'s program, \tool{MAW}, \tool{emMAW}, \tool{bwtman} and
\mm{} report the same number of MAWs of every length; on the two human chromosomes, Pinho et al.'s program, \tool{bwtman} and \mm{} report the same
number of MAWs. \tool{pMAW} agrees on three genomes; on
\emph{S.~cerevisiae} it reports 40,480,473 MAWs instead of 40,480,474, whatever the number of threads:
it misses four MAWs (\texttt{TCCCCGGGG} and its reverse complement, \texttt{CGTGTAAAAAAAAAG} and
\texttt{CGTGTAAAAAAAAAT}) and reports three words that occur in the genome (\texttt{CGTGTAAAAAAAAA},
\texttt{AGGTAAAAAAAAAAAAAAAAA} and \texttt{CGTAAAAAAAAAAAAAAAAAAA}), all checked by brute force. Above all,
\mm{} reproduces the 2015 distributions computed with Pinho et al.'s program: all 276 lengths of
\emph{E.~coli}, and all 7,822 lengths, from 11 to 1,475,836, of the 8,377,095,148 MAWs of the human
assembly GRCh38.p2 (Fig.~\ref{fig:dist}). The human genome with both strands is 6.1 billion bases, so
this also checks the 40-bit position path on real data.

\section{Scaling, and how it compares}

\subsection{Time}
The GPU's work grows linearly with the input: the radix sort takes about 2~ns per base on the human genome
and on the spruce alike (11.7~s for 6.1~G bases, 56~s for 24~G). Two parts grow faster. Each batch
rescans the whole text to pick its positions, so the extract step grows with the input times the number
of batches (2.1~s in 33 batches for human, 20~s in 131 for the spruce). And the deep stage, which runs on
the CPU, costs more per base on larger genomes, for two reasons: more of the positions lie in groups of
repeated 31-mers (22\% for human, 44\% for the spruce), and each of them is a random read into a packed
genome that no longer fits in the caches (1.5~GB for human, 6~GB for the spruce). It takes 25 core-ns per
base on human and 55 on the spruce, and it is the bottleneck on both: the CPU threads are busy 96 and
99\% of the time, while the GPU works 18~s of human's 29 and 88~s of the spruce's 221. Overall, \mm{}
takes about 2.4~ns per base (both strands) on chromosome~1, 4.7 on the human genome and 9 on the spruce.

The suffix-array programs grow somewhat faster than linearly as well (Fig.~\ref{fig:time}): from
chromosome~21 to chromosome~1, 5.7 times more bases, Pinho et al.'s time grows 7.6 times, \tool{MAW}'s 6.4
and \tool{bwtman}'s 7.0, against 3.5 for \mm{}, whose fixed costs dominate on small inputs. The
difference is in the constants. \tool{MAW} and Pinho et al.'s program are single-threaded and make
random accesses into arrays of several bytes per base; \tool{pMAW} parallelises only when its work
splitting finds independent parts, which fails as soon as the input contains an N (a fifth letter for
it), so on the human chromosomes it ran on one core; \tool{bwtman} uses about two cores; \tool{emMAW}
trades time for disk. \mm{} keeps 7 of the 10 cores and the GPU busy.

\subsection{Memory}
On genomes, \mm{} needs about half a byte per base, plus a fixed part; on texts made mostly of long
exact repeats it needs up to 9 bytes per base, since most of the text then reaches the tail (see Limitations). The packed text takes 0.25 bytes per base, the
segment bitmap 0.125, the repeat marks up to 0.125; the batch slots hold about a sixteenth of the positions (a
sixty-fourth in list mode), capped at 40\% of the GPU's recommended working set (about 8.5~GB on this laptop); and the tail needs 9 bytes per
residue base, after the batch buffers are freed. The peak resident memory was 1.4~GB for chromosome~1,
10.9~GB for the human genome and 17.5~GB for the spruce (Table~\ref{tab:mem}; macOS compresses part of
the rest, and the spruce's footprint peaked at 26~GB).

The earlier programs need far more, and the benchmark has to know how much before it runs them, since a
run that does not fit swaps for hours. We measured each program's peak memory on chromosome~21 and
expressed it in bytes per character of the indexed text, both strands together, with a margin: 28 for
Pinho et al.'s program (70 when it writes the words), 18 for \tool{MAW} (36 for its 64-bit build, needed
above $2^{31}$ characters), 16 for \tool{pMAW}, which has no 64-bit build, and 11 for \tool{bwtman};
\tool{emMAW} uses 8~GiB of RAM and about 20 bytes of temporary disk per character. These rules predict
the chromosome~1 peaks within the margin, and they give what the large genomes would need
(Table~\ref{tab:bench}): 66 to 216~GB for the human genome and 253 to 826~GB for the spruce, or 129 and
492~GB of disk for \tool{emMAW}.

\section{Benchmarks}

\subsection{Setup}
The genomes (Table~\ref{tab:genomes}) are those of our 2015 study where they could be identified,
including the human assembly we used then, GRCh38.p2, and the spruce assembly Pabies1.0, plus two human
chromosomes as intermediate sizes. Every program computed the MAWs of all lengths with the reverse
complement, all records together, under the protocol of the Methods. The main machine is a 2021 MacBook
Pro with an M1 Max and 32~GB. The same programs, built for each machine, also ran on the author's 2012
MacBook Pro (Core~i7-3720QM, four cores, 16~GB), the laptop of the 2015 study and comparable to the desktop
computers of the time; on a 2019 MacBook Pro (Core~i9-9980HK, eight cores, 64~GB); and on a 2023 MacBook
Pro (M3~Pro, 5 performance and 6 efficiency cores, 14-core GPU, 36~GB). \mm{} needs Apple Silicon, so it
ran only on the last two.

\begin{figure}[t]
  \tabcaption{tab:genomes}{Genomes. Mbp: ACGT bases on one strand; the programs index both strands.}
  \centering\footnotesize
  \begin{tabular}{@{}llr@{}}
    \toprule
    Genome & Source & Mbp \\
    \midrule
    HHV-5 (Merlin) & NC\_006273.2 & 0.24 \\
    \emph{E.~faecalis} V583 & GCF\_000007785.1 & 3.36 \\
    \emph{E.~coli} MG1655 & GCF\_000005845.2 & 4.64 \\
    \emph{S.~cerevisiae} R64 & GCF\_000146045.2 & 12.16 \\
    Human chr21 & UCSC hg38 & 40.09 \\
    Human chr1 & UCSC hg38 & 230.48 \\
    Human GRCh38.p2 & GCF\_000001405.28 & 3{,}061.46 \\
    Spruce Pabies1.0 & CNCB mirror & 12{,}018.81 \\
    \bottomrule
  \end{tabular}
\end{figure}

\begin{figure*}[t]
  \tabcaption{tab:bench}{Wall-clock seconds to compute all MAWs with the reverse complement on the 2021 laptop
    (M1 Max, 32~GB); medians of 6 runs (4 for chr1, 2 for the two largest genomes). Pinho et al.\ and \mm{}:
    counts mode, and in parentheses list mode, the words written to \texttt{/dev/null}. \tool{MAW} and
    \tool{emMAW} write the words, \tool{bwtman} counts them, \tool{pMAW} was built without its output
    stage. In brackets, for runs that do not fit on this laptop: the estimated memory (or disk) they
    need, or the indexing limit of the only build. Last column: the fastest earlier program over \mm{} in
    counts mode.}
  \centering\footnotesize
  \begin{tabular}{@{}lrrrrrrrr@{}}
    \toprule
    Genome & Mbp & Pinho et al.\ 2009 & \tool{MAW} 2014 & \tool{pMAW} 2016 & \tool{emMAW} 2017 & \tool{bwtman} 2017 & \mm{} 2026 & $\times$ \\
    \midrule
    \BenchMainTable
    \bottomrule
  \end{tabular}
\end{figure*}

\subsection{Results on the 2021 laptop}
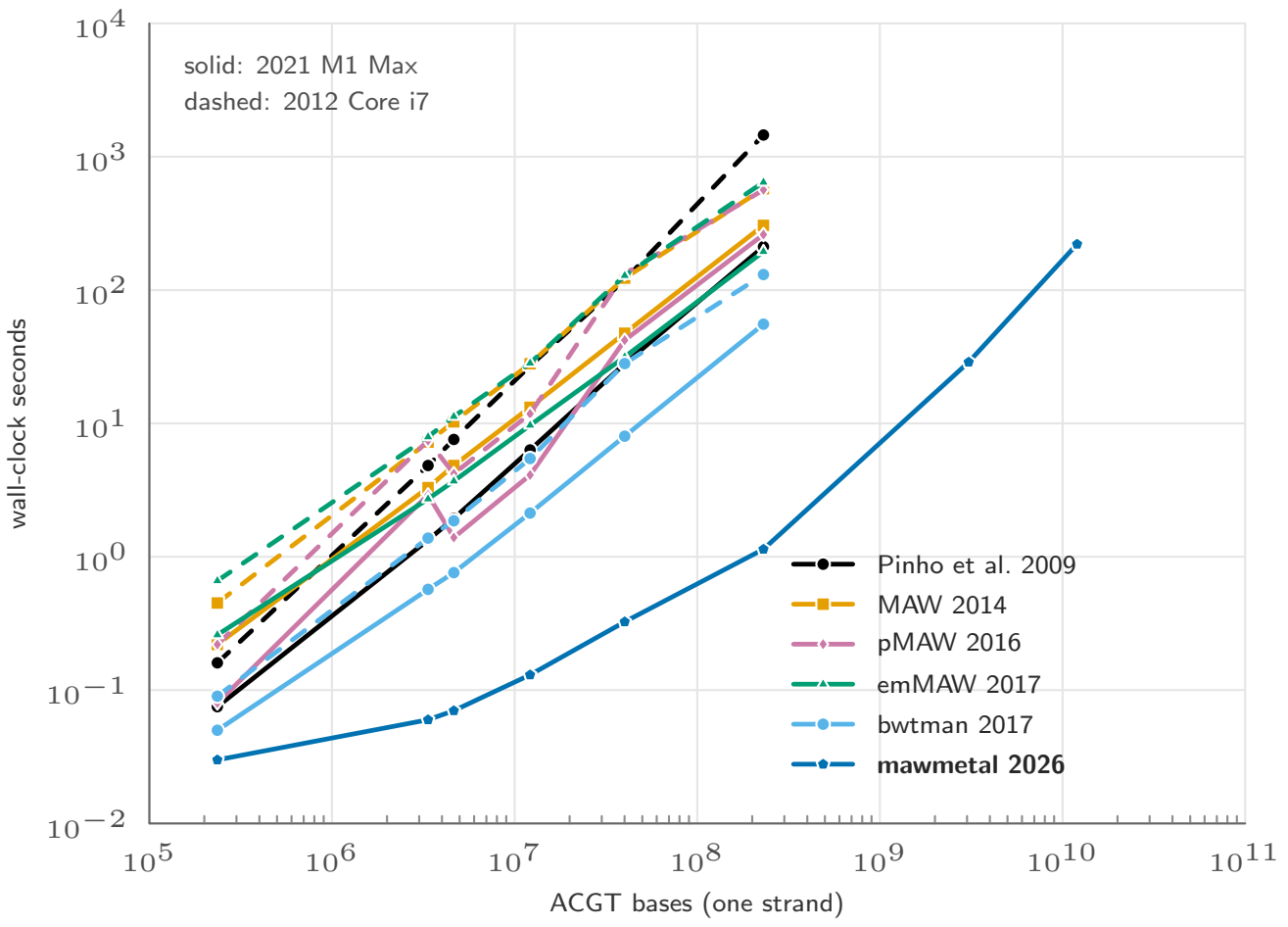
\begin{figure}[t]
  \centering
  \begin{tikzpicture}[font=\sffamily\scriptsize, line join=round, line cap=round]
\draw[black!10] (0.000,0) -- (0.000,5.400);
\node[below, black!70] at (0.000,-0.02) {$10^{5}$};
\draw[black!45] (0.371,0) -- +(0,0.05);
\draw[black!45] (0.588,0) -- +(0,0.05);
\draw[black!45] (0.743,0) -- +(0,0.05);
\draw[black!45] (0.862,0) -- +(0,0.05);
\draw[black!45] (0.960,0) -- +(0,0.05);
\draw[black!45] (1.042,0) -- +(0,0.05);
\draw[black!45] (1.114,0) -- +(0,0.05);
\draw[black!45] (1.177,0) -- +(0,0.05);
\draw[black!10] (1.233,0) -- (1.233,5.400);
\node[below, black!70] at (1.233,-0.02) {$10^{6}$};
\draw[black!45] (1.605,0) -- +(0,0.05);
\draw[black!45] (1.822,0) -- +(0,0.05);
\draw[black!45] (1.976,0) -- +(0,0.05);
\draw[black!45] (2.095,0) -- +(0,0.05);
\draw[black!45] (2.193,0) -- +(0,0.05);
\draw[black!45] (2.276,0) -- +(0,0.05);
\draw[black!45] (2.347,0) -- +(0,0.05);
\draw[black!45] (2.410,0) -- +(0,0.05);
\draw[black!10] (2.467,0) -- (2.467,5.400);
\node[below, black!70] at (2.467,-0.02) {$10^{7}$};
\draw[black!45] (2.838,0) -- +(0,0.05);
\draw[black!45] (3.055,0) -- +(0,0.05);
\draw[black!45] (3.209,0) -- +(0,0.05);
\draw[black!45] (3.329,0) -- +(0,0.05);
\draw[black!45] (3.426,0) -- +(0,0.05);
\draw[black!45] (3.509,0) -- +(0,0.05);
\draw[black!45] (3.580,0) -- +(0,0.05);
\draw[black!45] (3.644,0) -- +(0,0.05);
\draw[black!10] (3.700,0) -- (3.700,5.400);
\node[below, black!70] at (3.700,-0.02) {$10^{8}$};
\draw[black!45] (4.071,0) -- +(0,0.05);
\draw[black!45] (4.288,0) -- +(0,0.05);
\draw[black!45] (4.443,0) -- +(0,0.05);
\draw[black!45] (4.562,0) -- +(0,0.05);
\draw[black!45] (4.660,0) -- +(0,0.05);
\draw[black!45] (4.742,0) -- +(0,0.05);
\draw[black!45] (4.814,0) -- +(0,0.05);
\draw[black!45] (4.877,0) -- +(0,0.05);
\draw[black!10] (4.933,0) -- (4.933,5.400);
\node[below, black!70] at (4.933,-0.02) {$10^{9}$};
\draw[black!45] (5.305,0) -- +(0,0.05);
\draw[black!45] (5.522,0) -- +(0,0.05);
\draw[black!45] (5.676,0) -- +(0,0.05);
\draw[black!45] (5.795,0) -- +(0,0.05);
\draw[black!45] (5.893,0) -- +(0,0.05);
\draw[black!45] (5.976,0) -- +(0,0.05);
\draw[black!45] (6.047,0) -- +(0,0.05);
\draw[black!45] (6.110,0) -- +(0,0.05);
\draw[black!10] (6.167,0) -- (6.167,5.400);
\node[below, black!70] at (6.167,-0.02) {$10^{10}$};
\draw[black!45] (6.538,0) -- +(0,0.05);
\draw[black!45] (6.755,0) -- +(0,0.05);
\draw[black!45] (6.909,0) -- +(0,0.05);
\draw[black!45] (7.029,0) -- +(0,0.05);
\draw[black!45] (7.126,0) -- +(0,0.05);
\draw[black!45] (7.209,0) -- +(0,0.05);
\draw[black!45] (7.280,0) -- +(0,0.05);
\draw[black!45] (7.344,0) -- +(0,0.05);
\draw[black!10] (7.400,0) -- (7.400,5.400);
\node[below, black!70] at (7.400,-0.02) {$10^{11}$};
\draw[black!10] (0,0.000) -- (7.400,0.000);
\node[left, black!70] at (-0.02,0.000) {$10^{-2}$};
\draw[black!10] (0,0.900) -- (7.400,0.900);
\node[left, black!70] at (-0.02,0.900) {$10^{-1}$};
\draw[black!10] (0,1.800) -- (7.400,1.800);
\node[left, black!70] at (-0.02,1.800) {$10^{0}$};
\draw[black!10] (0,2.700) -- (7.400,2.700);
\node[left, black!70] at (-0.02,2.700) {$10^{1}$};
\draw[black!10] (0,3.600) -- (7.400,3.600);
\node[left, black!70] at (-0.02,3.600) {$10^{2}$};
\draw[black!10] (0,4.500) -- (7.400,4.500);
\node[left, black!70] at (-0.02,4.500) {$10^{3}$};
\draw[black!10] (0,5.400) -- (7.400,5.400);
\node[left, black!70] at (-0.02,5.400) {$10^{4}$};
\draw[black!60, thin] (0,5.400) -- (0,0) -- (7.400,0);
\node[black!80] at (3.700,-0.55) {ACGT bases (one strand)};
\node[rotate=90, black!80] at (-0.9,2.700) {wall-clock seconds};
\begin{scope}\clip (0,0) rectangle (7.400,5.400);
\draw[okBlack, line width=0.9pt, solid] plot coordinates {(0.459,0.788) (1.882,1.913) (2.056,2.059) (2.571,2.521) (3.210,3.103) (4.147,3.893)};
\draw[okBlack, line width=0.4pt] plot[only marks, mark=*, mark size=1.3pt, mark options={draw=white, fill=okBlack}] coordinates {(0.459,0.788) (1.882,1.913) (2.056,2.059) (2.571,2.521) (3.210,3.103) (4.147,3.893)};
\draw[okBlack, line width=0.9pt, dashed] plot coordinates {(0.459,1.084) (1.882,2.416) (2.056,2.592) (2.571,3.089) (3.210,3.669) (4.147,4.647)};
\draw[okBlack, line width=0.4pt] plot[only marks, mark=*, mark size=1.3pt, mark options={draw=white, fill=okBlack}] coordinates {(0.459,1.084) (1.882,2.416) (2.056,2.592) (2.571,3.089) (3.210,3.669) (4.147,4.647)};
\draw[okOrange, line width=0.9pt, solid] plot coordinates {(0.459,1.208) (1.882,2.267) (2.056,2.416) (2.571,2.809) (3.210,3.309) (4.147,4.037)};
\draw[okOrange, line width=0.4pt] plot[only marks, mark=square*, mark size=1.3pt, mark options={draw=white, fill=okOrange}] coordinates {(0.459,1.208) (1.882,2.267) (2.056,2.416) (2.571,2.809) (3.210,3.309) (4.147,4.037)};
\draw[okOrange, line width=0.9pt, dashed] plot coordinates {(0.459,1.488) (1.882,2.574) (2.056,2.711) (2.571,3.103) (3.210,3.681) (4.147,4.284)};
\draw[okOrange, line width=0.4pt] plot[only marks, mark=square*, mark size=1.3pt, mark options={draw=white, fill=okOrange}] coordinates {(0.459,1.488) (1.882,2.574) (2.056,2.711) (2.571,3.103) (3.210,3.681) (4.147,4.284)};
\draw[okPurple, line width=0.9pt, solid] plot coordinates {(0.459,0.813) (1.882,2.218) (2.056,1.930) (2.571,2.352) (3.210,3.262) (4.147,3.975)};
\draw[okPurple, line width=0.4pt] plot[only marks, mark=diamond*, mark size=1.3pt, mark options={draw=white, fill=okPurple}] coordinates {(0.459,0.813) (1.882,2.218) (2.056,1.930) (2.571,2.352) (3.210,3.262) (4.147,3.975)};
\draw[okPurple, line width=0.9pt, dashed] plot coordinates {(0.459,1.208) (1.882,2.584) (2.056,2.361) (2.571,2.765) (3.210,3.710) (4.147,4.277)};
\draw[okPurple, line width=0.4pt] plot[only marks, mark=diamond*, mark size=1.3pt, mark options={draw=white, fill=okPurple}] coordinates {(0.459,1.208) (1.882,2.584) (2.056,2.361) (2.571,2.765) (3.210,3.710) (4.147,4.277)};
\draw[okGreen, line width=0.9pt, solid] plot coordinates {(0.459,1.273) (1.882,2.190) (2.056,2.311) (2.571,2.688) (3.210,3.149) (4.147,3.860)};
\draw[okGreen, line width=0.4pt] plot[only marks, mark=triangle*, mark size=1.3pt, mark options={draw=white, fill=okGreen}] coordinates {(0.459,1.273) (1.882,2.190) (2.056,2.311) (2.571,2.688) (3.210,3.149) (4.147,3.860)};
\draw[okGreen, line width=0.9pt, dashed] plot coordinates {(0.459,1.638) (1.882,2.609) (2.056,2.747) (2.571,3.107) (3.210,3.698) (4.147,4.328)};
\draw[okGreen, line width=0.4pt] plot[only marks, mark=triangle*, mark size=1.3pt, mark options={draw=white, fill=okGreen}] coordinates {(0.459,1.638) (1.882,2.609) (2.056,2.747) (2.571,3.107) (3.210,3.698) (4.147,4.328)};
\draw[okSky, line width=0.9pt, solid] plot coordinates {(0.459,0.629) (1.882,1.580) (2.056,1.693) (2.571,2.095) (3.210,2.614) (4.147,3.370)};
\draw[okSky, line width=0.4pt] plot[only marks, mark=*, mark size=1.3pt, mark options={draw=white, fill=okSky}] coordinates {(0.459,0.629) (1.882,1.580) (2.056,1.693) (2.571,2.095) (3.210,2.614) (4.147,3.370)};
\draw[okSky, line width=0.9pt, dashed] plot coordinates {(0.459,0.859) (1.882,1.926) (2.056,2.043) (2.571,2.463) (3.210,3.104) (4.147,3.705)};
\draw[okSky, line width=0.4pt] plot[only marks, mark=*, mark size=1.3pt, mark options={draw=white, fill=okSky}] coordinates {(0.459,0.859) (1.882,1.926) (2.056,2.043) (2.571,2.463) (3.210,3.104) (4.147,3.705)};
\draw[okBlue, line width=0.9pt, solid] plot coordinates {(0.459,0.429) (1.882,0.700) (2.056,0.761) (2.571,1.003) (3.210,1.361) (4.147,1.851) (5.533,3.114) (6.265,3.910)};
\draw[okBlue, line width=0.4pt] plot[only marks, mark=pentagon*, mark size=1.3pt, mark options={draw=white, fill=okBlue}] coordinates {(0.459,0.429) (1.882,0.700) (2.056,0.761) (2.571,1.003) (3.210,1.361) (4.147,1.851) (5.533,3.114) (6.265,3.910)};
\end{scope}
\draw[okBlack, line width=0.9pt, solid] (4.35,1.750) -- +(0.4,0);\draw[okBlack] plot[only marks, mark=*, mark size=1.3pt, mark options={draw=white, fill=okBlack}] coordinates {(4.55,1.750)};\node[right] at (4.80,1.750) {Pinho et al.\ 2009};
\draw[okOrange, line width=0.9pt, solid] (4.35,1.480) -- +(0.4,0);\draw[okOrange] plot[only marks, mark=square*, mark size=1.3pt, mark options={draw=white, fill=okOrange}] coordinates {(4.55,1.480)};\node[right] at (4.80,1.480) {MAW 2014};
\draw[okPurple, line width=0.9pt, solid] (4.35,1.210) -- +(0.4,0);\draw[okPurple] plot[only marks, mark=diamond*, mark size=1.3pt, mark options={draw=white, fill=okPurple}] coordinates {(4.55,1.210)};\node[right] at (4.80,1.210) {pMAW 2016};
\draw[okGreen, line width=0.9pt, solid] (4.35,0.940) -- +(0.4,0);\draw[okGreen] plot[only marks, mark=triangle*, mark size=1.3pt, mark options={draw=white, fill=okGreen}] coordinates {(4.55,0.940)};\node[right] at (4.80,0.940) {emMAW 2017};
\draw[okSky, line width=0.9pt, solid] (4.35,0.670) -- +(0.4,0);\draw[okSky] plot[only marks, mark=*, mark size=1.3pt, mark options={draw=white, fill=okSky}] coordinates {(4.55,0.670)};\node[right] at (4.80,0.670) {bwtman 2017};
\draw[okBlue, line width=0.9pt, solid] (4.35,0.400) -- +(0.4,0);\draw[okBlue] plot[only marks, mark=pentagon*, mark size=1.3pt, mark options={draw=white, fill=okBlue}] coordinates {(4.55,0.400)};\node[right] at (4.80,0.400) {\textbf{mawmetal 2026}};
\node[anchor=north west, black!70, align=left] at (0.12,5.3) {solid: 2021 M1 Max\\dashed: 2012 Core i7};
\end{tikzpicture}
  \caption{Wall-clock time against genome size: the 2021 M1 Max laptop (solid) and the 2012 MacBook Pro
    (dashed), counts mode where a program has one. Only \mm{} goes beyond human chromosome~1; on the
    2012 laptop only the earlier programs run.\label{fig:time}}
\end{figure}

All programs agree wherever they ran, except \tool{pMAW} on yeast (see Validation). Table~\ref{tab:bench}
and Fig.~\ref{fig:time} give the times. On every genome \mm{} is the fastest, and the gap widens with the
genome: \BEcoliSpeedup{} times the fastest earlier program on \emph{E.~coli}, \BChrTwentyOneSpeedup{} on
chromosome~21 and \BChrOneSpeedup{} on chromosome~1, where it takes \BChrOneMm~s against
\BChrOneBwt~s for \tool{bwtman}, \BChrOneEmmaw~s for \tool{emMAW}, \BChrOnePinho~s for Pinho et al.'s
program, \BChrOnePmaw~s for \tool{pMAW} and \BChrOneMaw~s for \tool{MAW}. In list mode, against the
programs that write words, the factor is larger still: \BChrOneMmList~s against \BChrOneEmmaw~s for
\tool{emMAW} and \BChrOneMaw~s for \tool{MAW}.

Beyond chromosome~1, \mm{} is alone. The human genome, 6.1~billion bases with both strands, takes it
\BHumanMm~s in counts mode and \BHumanMmList~s in list mode, with a peak of \BHumanMmMem~GB; the earlier
programs would need between 66 and 216~GB of memory, or 129~GB of disk. The spruce, 24~billion bases in
10~million scaffolds, takes \BSpruceMm~s, under four minutes, with a peak resident memory of
\BSpruceMmMem~GB.

This deserves to be said plainly. In 2015 the spruce took several days on a supercomputer with a
terabyte of memory. In 2026, code written by a language model computes the same MAWs on a laptop with
32~GB, in under four minutes. It is the only program that can process such a genome on that laptop: the
others would need 253 to 826~GB of memory in counts mode (1.6~TB for Pinho et al.'s list mode), cannot
index it (\tool{pMAW}), or, for \tool{emMAW}, about 500~GB of free disk (estimated from its rule of 20
bytes per character) where the laptop had 41~GB, so it was not attempted, and the fastest of them, on the
largest genome they all fit, is \BChrOneSpeedup{} times slower.

\begin{figure*}[t]
  \tabcaption{tab:mem}{Peak resident memory (GB) on the 2021 laptop; in brackets, the estimate for runs that
    do not fit (see Scaling).}
  \centering\footnotesize
  \begin{tabular}{@{}lrrrrrrrr@{}}
    \toprule
    Genome & Pinho & Pinho list & \tool{MAW} & \tool{pMAW} & \tool{emMAW} & \tool{bwtman} & \mm{} & \mm{} list \\
    \midrule
    \BenchMemTable
    \bottomrule
  \end{tabular}
\end{figure*}

\subsection{Ten years of hardware}
Table~\ref{tab:machines} runs the same programs on the four laptops. From the 2012 laptop to the 2021
one, the earlier programs became 1.9 to 6.9 times faster on chromosome~1, and 2.6 to 4.3 times on
chromosome~21; Pinho et al.'s program gains the most, though on chromosome~1 the 2012 laptop was close to
its 16~GB and its two runs took 946 and 1,967~s. The 2019 Intel laptop sits in between, and the 2023 M3 Pro, with fewer performance cores and GPU cores, runs most earlier programs 4 to 15\% faster
than the M1 Max and \mm{} 1.4 to 1.7 times slower.

Put together, the numbers separate three sources of speed. Ten years of \emph{hardware}, with its compilers and
operating systems, gave a factor of 2 to 7 to unchanged programs. Better \emph{algorithms} gave less than one might expect: the
linear-time \tool{MAW} is slower than the 2009 program, quadratic in the worst case, on the 2021 laptop
(\BChrOneMaw~s against \BChrOnePinho~s on chromosome~1), though not on the 2012 one, and the best of the
compressed-space methods, \tool{bwtman}, is the fastest earlier program by a factor of 3.5. Fitting the
algorithm to the \emph{architecture} gave the rest, in two parts that this work did not measure separately:
tiers that keep the suffix array for the 5 to 7\% of bases in long repeats, and a GPU, idle under the earlier
programs, that sorts the rest: \mm{} on the 2021 laptop is \BChrOneSpeedup{} times
faster than \tool{bwtman} on the same laptop, and \BChrOneTotalGain{} times faster than the 2015 software on
the 2015 laptop. Which program is fastest depends on the machine, and an algorithm written for a
machine beats one that is only asymptotically better.

\begin{figure*}[t]
  \tabcaption{tab:machines}{The same programs on four laptops: wall-clock seconds, counts mode, and the
    2012-to-2021 ratio.}
  \centering\footnotesize
  \begin{tabular}{@{}llrrrrr@{}}
    \toprule
    & & 2012 & 2019 & 2021 & 2023 & \\
    & & i7 & i9 & M1 Max & M3 Pro & ratio \\
    \midrule
    \BenchMachineTable
    \bottomrule
  \end{tabular}
\end{figure*}

\section{Limitations}
\mm{} runs only on Apple Silicon. The design would carry over to other GPUs with unified memory, and in
part to discrete ones, but the code would not.

Its size is limited. The GPU kernels index the packed text with 32-bit word indices, which caps a run at
$2^{36}$ bases, 68.7~Gbp including the reverse complement, and the whole text must fit in one Metal
buffer, at most about 20~GB on this laptop (80~Gbp). The largest genome measured, that of the fork fern
\emph{Tmesipteris oblanceolata}, 160~Gbp~\cite{Fernandez2024}, has no assembly; the largest assembled one
near it, the mistletoe \emph{Viscum album} (94~Gbp, GCA\_963277665.1), would exceed both limits as one
unit, though not chromosome by chromosome. Lifting them needs 64-bit word indices and a text split
across buffers, and then about 0.5 bytes per base of memory, so a computer with several hundred
gigabytes.

Its cost grows with the repeat content. On inputs made mostly of long exact repeats, nearly every
position reaches the tail: two copies of a 2.2~Gbp genome, a test written during development, need about
100~GB for their suffix array and do not fit in 32~GB. For collections of near-identical genomes,
compressed-space methods~\cite{Nishimoto2021,Inenaga2024} are the right tools. Per-record mode over
millions of small scaffolds pays the GPU's setup for each record; \texttt{--joint} avoids it.

Finally, \mm{} is new code, written by an LLM, and trusted because of its tests and its agreement with
independent programs and with our 2015 results, not because of years of use.

\section{How the work was done}

\subsection{Setup}
All of the work was done by Claude~\cite{ClaudeCode}, a large language model of Anthropic: first in a
conversation in the claude.ai web application (model not recorded), then through Claude Code, an agent
that reads and writes files and runs commands on the user's computer, with the model Claude Opus~5.5,
effort \texttt{xhigh} and adaptive thinking, in a permission mode where the agent runs commands without
asking for each. It ran on the author's Claude Pro subscription, \$20 per month. The author wrote no code
and no prose. The author's part was to choose the problem, ask questions, object, set priorities,
decide trade-offs when asked, supply the 2009 software, the 2015 results and the old laptop, and take
responsibility for the content, as arXiv requires~\cite{arXivAI}.

\subsection{Process}
The work took about five days end to end, in the author's spare time, in \LLMSessions{} Claude Code
sessions, some running in
parallel (Table~\ref{tab:usage}). Table~\ref{tab:prompts} quotes the prompts that shaped it; the Appendix
lists all \LLMPrompts{}, verbatim. The pattern was the same throughout. The author stated a goal in a
sentence or two, often loosely (``We want this code to be absolutely bonker fast''); the agent explored
the code and the machine, measured, proposed a plan with the trade-offs it saw, carried it out once the
author agreed, and reported what it had measured. Several of the decisive turns were the author's: the
objection that the longest MAWs are very long, which led to the tiers; the remark that only the
distribution is needed, which made counts mode the default; the observation that 40 bits of position
are enough; the request to focus on large genomes, which reversed several optimizations tuned on
chromosome~1. The agent's own initiatives were as important: it proposed the deep stage, built every
earlier program from source and fixed their portability problems without touching their algorithms,
found that \tool{pMAW} misses MAWs and why it runs on one core, located the author's copy of the 2009
software on the author's disk, and recognized the 2015 human assembly from the base count in our old logs.

\begin{figure*}[t]
  \tabcaption{tab:prompts}{The prompts that shaped the work, verbatim (shortened where marked). All of them are
    in the Appendix. S$n$: Claude Code session (Table~\ref{tab:usage}).}
  \centering\footnotesize
  \begin{tabular}{@{}p{0.07\textwidth}p{0.55\textwidth}p{0.34\textwidth}@{}}
    \toprule
    Where & Prompt & Outcome \\
    \midrule
    claude.ai & ``What's the state of the art when it comes to code for finding these nowadays ?'' & A review of \tool{MAW}, \tool{emMAW}, BWT methods. \\
    claude.ai & ``Can any of these method can be made way faster by using metal ?'' & GPU bitmaps for short MAWs. \\
    claude.ai & ``But the longest MAW can be very very long...'' & The tiered design and the exactness argument. \\
    S1 & ``Transfering a conversation from claude online:'' [the chat, pasted] & First version, tests, validation (one turn, 20.7~h elapsed). \\
    S1 & ``The Spruce genome took days then. How long would it take with this code ? (estimate)'' & An estimate, and the idea of the deep stage. \\
    S1 & ``Can you build that new stage ?'' & The deep stage. \\
    S1 & ``we don't need to actually output the maw list. Just compute them, count them and discard. The distribution is enough.'' & Counts mode by default; \texttt{--list}. \\
    S1 & ``40 or 48 bits is plenty'' & 40-bit positions. \\
    S4 & ``We want this code to be absolutely bonker fast. [\dots] Think about memory management, cache optimizition, avoiding cache misses [\dots]'' & Parallel input, GPU tier 1, 3 slots: chr1 3$\times$ faster, 3$\times$ less memory. \\
    S5 & ``Get the full spruce genome and add it to the benchmark. [\dots] The benchmark should include [\dots] any other code publsihed in between'' & The benchmark suite, with \tool{bwtman}. \\
    S9 & ``We mostly need optimizations for large genomes! That's where it matters.'' & Retuning for human and spruce; the spruce fits. \\
    S10 & ``Look at libsais. [\dots] Can we make them faster specifically on Apple silicon by using metal ?'' & No; the deep stage is the lever. Tail memory 13$\to$9 B/base. \\
    S7 & ``Any chances to run the genome of New Caledonian fork fern (Tmesipteris oblanceolata) with mawmetal ?'' & No assembly exists; the 68.7~Gbp limit. \\
    \bottomrule
  \end{tabular}
\end{figure*}

\subsection{What it cost}
Claude Code keeps a transcript of every session, with the token usage of every model call, from which
Table~\ref{tab:usage} is computed. Over \LLMSessions{} sessions the author typed \LLMPrompts{} prompts,
\LLMPromptWords{} words in all, plus a pasted chat. The model made
\LLMApiCalls{} calls and \LLMToolCalls{} tool calls, mostly shell commands; it read \LLMInputMTok{} million
input tokens, \LLMCacheReadMTok{} million of them from the prompt cache, and wrote \LLMOutputMTok{} million
output tokens, \LLMThinkingTokens{} of them reasoning. At Anthropic's list prices these calls would cost
\$\LLMCostUSD{} through the API; on the subscription they cost \$20. The agents were busy for
\LLMWorkingHours{} hours in all, counting gaps of under ten minutes between events; the rest was waiting
for benchmarks, a laptop asleep, or the author away. The author's own time, reading the reports and
writing the prompts, was about 4 to 8 hours over those five days, all of it in spare time.

\begin{figure*}[t]
  \tabcaption{tab:usage}{Claude Code sessions. Start: hours after the first prompt. Hours: agent working time; input in
    millions and output in thousands of tokens (input includes cache reads and writes); cost at API list
    prices. }
  \centering\footnotesize
  \begin{tabular}{@{}llp{0.38\textwidth}rrrrrr@{}}
    \toprule
    & Start & Purpose & Prompts & Hours & Calls & Input M & Output k & Cost \$ \\
    \midrule
    \LLMSessionTable
    \bottomrule
  \end{tabular}
\end{figure*}

\subsection{What went wrong}
The transcripts show the agents working around their environment as much as around their own bugs. The
disk of the development laptop was full, and \tool{MAW}'s 5.5~GB output silently truncated \mm{}'s own;
the agent made \mm{} fail loudly on write errors and compared large outputs through an
order-independent hash. Memory was exhausted for hours by unrelated applications, which stalled runs;
the agent diagnosed it from CPU times and memory statistics. The compiler toolchain lacked its C++
headers, the package manager was the Intel one, and the Python of the shell ran under Rosetta, which
silently made the benchmark time Intel binaries until an agent noticed. The agents made mistakes of their
own and caught most of them: a change to the sort placed keys in the wrong region, which the randomized
test flagged within two minutes; one diagnosis of a slow start was wrong, and was corrected by the next
measurement; an optimization that helped the human genome made the spruce 47\% slower and was taken
out. One was not caught in time: the paper-writing session ran the development session's test of two
identical 2.2~Gbp copies without noticing that it sends everything to the tail, and the laptop reached
23~GB of swap before the agent stopped it.

\subsection{What it means for scientists}
None of the techniques in \mm{} is beyond a good programmer. What is new is the cost: a few days, a few
dozen prompts and a \$20 subscription, for work that we would expect to take a specialist months and that nobody
would have funded for a one-off computation. For a scientist with a slow program, the lesson is practical.
Ask for a review of the code ``from a performance lens'', for a plan, for measurements, and for the
trade-offs; insist on a benchmark against every alternative and on independent checks of the output. The
LLM does the tedious parts that used to make such work impractical, from building other people's code
to running a benchmark on four machines, and it does them without being asked twice. The judgment of
what matters, and the responsibility for the result, stay with the scientist.

\section{Conclusions}
Ten years after our study of minimal absent words, the computation that needed a 1-terabyte
supercomputer and several days runs on a 32~GB laptop in under four minutes, and the human distribution
we published is reproduced exactly in half a minute. Three sources of speed can be separated. Ten years
of hardware gave a factor of 2 to 7. Better algorithms, linear-time and compressed-space, gave less than
their asymptotics promise, and on today's laptop the 2009 program beats its linear-time successor.
Fitting the algorithm to the architecture gave a
factor of 49 over the best earlier program. It was not a port of an earlier algorithm but a new division of the work, by the
structure of the input and the strengths of each processor, and it made genomes of tens of gigabases
possible at all.

That last kind of work used to be impractical: it is specialized, specific to one machine, and costly.
With a language model it took days. Claude designed the algorithm, implemented it, optimized it,
validated it against every program we could find and against our old results, benchmarked them all on
four machines, and wrote this paper. We expect the same to hold far beyond MAWs: many scientific programs
run on hardware they were not written for, and the cost of fitting them to it has just dropped by orders
of magnitude.

\begin{materials}

  \section{Data}
  Human herpesvirus 5 (NC\_006273.2), \emph{E.~faecalis} V583 (GCF\_000007785.1, chromosome and three
  plasmids), \emph{E.~coli} K-12 MG1655 (GCF\_000005845.2), \emph{S.~cerevisiae} S288C (GCF\_000146045.2)
  and the human assembly GRCh38.p2 (GCF\_000001405.28, 486 records including alternate loci and patches)
  are from NCBI; human chromosomes 1 and 21 from UCSC hg38; the Norway spruce assembly Pabies1.0~\cite{Nystedt2013} from a mirror at the China National Center
  for Bioinformation. URLs, sizes and sha256 checksums are in
  \texttt{bench/genomes.tsv}. GRCh38.p2 is the release of our 2015 study: its 3,061,458,595 ACGT bases
  match our 2015 logs. The 2015 distributions come from the author's files of that study.

  \section{Software}
  Pinho et al.'s program is the 2009 release, distributed here with the kind permission of A.~J.~Pinho,
  with one declaration fixed (\texttt{gzFile *fp} to \texttt{gzFile fp}) for current zlib. \tool{MAW},
  \tool{pMAW} and \tool{emMAW} are from the \tool{MAW} suite (github.com/solonas13/maw, commit
  d69fa88), with the sdsl-lite, libdivsufsort~2.0.1, pSAscan and EM-SparsePhi it ships; \tool{bwtman} is
  from github.com/fcunial/portfolio (commit 4f37fa6). All were built with Apple clang (or the Command Line
  Tools of each machine) and an OpenMP runtime built from LLVM~19.1.7; the portability fixes, none of them
  in an algorithm, are listed in \texttt{bench/build\_tools.sh}. \mm{} uses libsais~\cite{libsais} (commit
  c284ba8) and libdeflate~\cite{libdeflate}.

  \section{Benchmark protocol}
  Every program computed the MAWs of all lengths (bound 5,000,000, as in 2015, which no MAW reaches:
  the longest, in the human genome, has 1,475,836 bases) of the genome with its reverse complement, all records together. Pinho et al.'s program, \tool{bwtman} and \mm{} cut the
  sequence at every letter other than ACGT. \tool{MAW}, \tool{pMAW} and \tool{emMAW} take one record
  and read N as a fifth letter, so they received the genome as one record in which every record boundary
  and every run of other letters is a single N, and their words containing N were not counted; this leaves
  the MAWs over ACGT unchanged. \tool{pMAW} and \tool{bwtman} were given all cores, \tool{emMAW} 8~GiB of
  RAM. Output was discarded in each program's cheapest way (Table~\ref{tab:bench}). Each program ran three
  times per genome (twice for chr1, once for the two largest genomes) in each of two benchmark runs on the
  M1 Max, the programs taking turns, under \texttt{caffeinate}, on mains power, with the input in the page
  cache; one run on each of the other machines. Wall-clock time, CPU time and peak resident memory come
  from \texttt{/usr/bin/time}; the machine's load, free memory and swap before each run are recorded. A
  run is skipped when its estimated memory or disk exceeds the machine's (Scaling), and killed if swap
  grows by more than 24~GB. Separate runs counted the MAWs of every length for every program. Everything
  is in \texttt{bench/}, one command (\texttt{bench/run.sh}) with a profile for the 16~GB laptop, and each
  run writes one result file.

  \section{LLM accounting}
  The numbers of \emph{How the work was done} are computed by \texttt{paper/tools/llm\_usage.py} from the
  Claude Code transcripts (Claude Code 2.1.308, model \texttt{claude-opus-5-5}). Prices are Anthropic's
  list prices for that model, per million tokens: \$4 for input, \$8 for one-hour cache writes, \$0.20 for
  cache reads, \$20 for output. The claude.ai chat is not included: its transcript was not recorded.

  \section{Availability}
  The code of \mm{}, the benchmark and this paper are released under the MIT license.
  The source code is at \url{https://github.com/tralhavet/mawmetal}.
\end{materials}

\begin{acknowledgments}
  The author thanks Armando J.~Pinho, whose 2009 software made our 2015 study possible, for permission
  to distribute it with this work, so that the comparison can be reproduced; Erik Aurell and Hai-Jun
  Zhou, coauthors of the 2015 study; and the authors of \tool{MAW}, \tool{pMAW}, \tool{emMAW},
  \tool{bwtman} and libsais for releasing their code.
\end{acknowledgments}

\section{Appendix: every prompt}
{\footnotesize\raggedright
% Tight enough that the appendix ends on its page: a few lines carried over overflow the class's last page.
\setlength{\smallskipamount}{0pt}\setlength{\baselineskip}{0.97\baselineskip}%
Every prompt the author typed in Claude Code, verbatim, by session (Table~\ref{tab:usage}), with its time in
hours after the first one; $^q$: typed while
the agent was working, and delivered to it mid-turn. Pasted texts are replaced by their length. Not
shown: the three questions of the claude.ai chat (Table~\ref{tab:prompts}), which are inside the first
pasted text.\par\smallskip
% Generated by tools/llm_usage.py; do not edit.
\par\smallskip\noindent\textbf{S1. Algorithm: first version, then sort, overlap, deep stage} (from t+0.0\,h)\par
\noindent{\ttfamily t+0.0\,h}\enspace Transfering a conversation from claude online: [pasted text, 9,708 characters]\par
\noindent{\ttfamily t+49.5\,h}\enspace Do all that: A multithreaded libsais (needs an arm64 OpenMP). Overlapping the GPU sort of one batch with the CPU scan of the previous one. A faster sort.\par
\noindent{\ttfamily t+49.5\,h}$^q$\enspace I wrote that paper a decade ago. \url{https://arxiv.org/abs/1509.05188} The Spruce genome took days then. How long would it take with this code ? (estimate) How much memory/disk space would I need for that ?\par
\noindent{\ttfamily t+49.7\,h}$^q$\enspace Can you build that new stage ? How would that look like ,\par
\noindent{\ttfamily t+49.7\,h}$^q$\enspace we don't need to actually output the maw list. Just compute them, count them and discard. The distribution is enough. (Now, some people may want the list, so a comand line argument for setting the output is best)\par
\noindent{\ttfamily t+50.4\,h}$^q$\enspace You don't really need 64 bits.... Largest genome is 40 G base pairs or so.... 40 or 48 bits is plenty\par
\par\smallskip\noindent\textbf{S2. Finding Pinho et al.'s 2009 code} (from t+49.8\,h)\par
\noindent{\ttfamily t+49.8\,h}\enspace Pinho, A, Ferreira, P, Garcia, S, \& Rodrigues, J. (2009) On finding minimal absent words. BMC Bioinformatics 10, 137. Can you find the code for this somewhere ?\par
\noindent{\ttfamily t+49.9\,h}$^q$\enspace might even be on this machine....\par
\noindent{\ttfamily t+49.9\,h}\enspace I found independently! \textasciitilde{}/DOwnloads/maws is probably the original code by Pinho et all Then the code attached is probably the modified version I used for the paper. Add them to this project :)\par
\par\smallskip\noindent\textbf{S3. This paper} (from t+50.1\,h)\par
\noindent{\ttfamily t+50.1\,h}\enspace Let's start writing a new paper in the same style as my 2015 paper \url{https://arxiv.org/pdf/1509.05188} The paper will start by reviewing what algorithms have been created after the Pinho et al. It will then go on and describe the algorithm we have just made in detail, and explain its advantages and drawbacks, comparing fairly to all others. We'll also have a section where we benchmark all of the algorithms on some of the genomes in my 2015 paper. The bulk of the comparison will be on this Machine. I happen to still own the machine used for most of the computations of the 2015 paper, so we can also run a lighter benchmark comparing this machine and the other. Then it will mention that it's 2026 and we now have Claude and LLM.... And we'll disclose that the algorithm as well as this paper was written entirely by Claude (on the 20\$/month contract). If we can, we can describe the procedure we used (I think you have all the conversations and prompts, so you'll be able to check). We can try to estimate how much work we put in and how many tokens. The punchline will probably be (but need to see the result of the benchmark) that 1. Computers haven't changed that much 2. Algorithms make all the diffference 3. Inventing new algorithm, running such benchmarks and producing papers about it is now orders of magnitude easier than before with LLMs. In fact, this work would never have been attempted at all without AI.\par
\noindent{\ttfamily t+50.3\,h}$^q$\enspace The journal version is \url{https://iopscience.iop.org/article/10.1088/1478-3975/13/2/026004} It's the same\par
\noindent{\ttfamily t+61.1\,h}\enspace continue\par
\noindent{\ttfamily t+68.9\,h}\enspace Notes for the next version of the paper (take notes, don't act now): - You'll need to explain a bit more the apple silicon architeture, what's special, and what part of the algorithm are tuned to it. - You probably took inspiration from existing algorithms in the litterature, at least by part, to create mawmetal --- I'd like to make it clear which ideas were taken from where, and what is truly novelty created by LLM. - I suggest you try to explain the algorithm better, perhaps with diagrams, etc.... - In another conversation titled "New Caledonian fork fern (Tmesipteris oblanceolata)", we discoevered a limitation to mawmetal that is worth discussion in the paper\par
\noindent{\ttfamily t+70.8\,h}\enspace In the description of the algorithm, you'll also describe how it scales in speed and memory usage (maybe separating CPU vs GPU scaling), and compare to other algorithms.\par
\noindent{\ttfamily t+70.8\,h}\enspace You'll insist that it's the only algo that can do the spruce genome on the laptop. You'll also explain how you estimate the memory requirements of others (can be part of the scaling discussion).\par
\noindent{\ttfamily t+70.8\,h}\enspace you'll clearly explain the difference between list and non list modes\par
\noindent{\ttfamily t+70.9\,h}\enspace You'll add a punchline: The spruce took several days on a super computer at PDC KTH with 1TB ram. Code created by LLM runs it on a laptop with 32 GB ram. It's the only code that can process such a genome on this laptop. No other code found in the literature come close (we'll need to show numbers in the benchmark section). Let that sink!\par
\noindent{\ttfamily t+72.8\,h}\enspace Make it more obvious from the text and the abstract that this is as much a new result as an experiment in using LLM to design and optimize new algorithms. The use of Claude should be core of this work, not a footnote. Ultimately - beyond the new results and the review value of the article - I want scientists doing computations to understand that Claude is a tool they can use for massive improving scientific code at low cost and low amount of efforts\par
\noindent{\ttfamily t+86.7\,h}\enspace I've contated Pinho and he's fine sharing his the original code from 2009 as part of this project. Note that and remember to emphasize it on the readme and in the aknowledgements.\par
\noindent{\ttfamily t+99.4\,h}\enspace You'll change the punchline of the paper to be more towards benefits of using llms for optimizationa and for science in general. You'll point out that optimizing for the architecture beats better algorithms, and that while it was impractical before, with LLMs it's dead easy.\par
\noindent{\ttfamily t+99.5\,h}\enspace Correction: I don't have the machines that were used for the computation itself anymore. I do have the laptop that I used at the time - Macbook pro Retina 2012 2.6Ghz Core i7 (quad) with 16GB of ram. It was probably comparable to the desktop machines available at the time.\par
\noindent{\ttfamily t+110.3\,h}\enspace and now results from the benchmark are in Take all the notes I dumped onto you previously, these benchmark results and create the next version of the paper based on all that\par
\noindent{\ttfamily t+110.5\,h}$^q$\enspace Lift the thing about the LLM. Clearly say in the abstract already that we disigned the algorithm using claude, feeding it with literature and got a result that literaly obliterates anything out there\par
\par\smallskip\noindent\textbf{S4. Performance review of mawmetal} (from t+61.3\,h)\par
\noindent{\ttfamily t+61.3\,h}\enspace Look at the code of maw\_metal. Review it from code execution performance lens. We want this code to be absolutely bonker fast. Make sure we are applying all possible optimizations. Plan for implementing any optimization you can still find. Think about memory management, cache optimizition, avoiding cache misses, aligning stuff in memory for speed, loop unfolding, etc... All those tricks --- apply whenever possible.\par
\noindent{\ttfamily t+68.0\,h}\enspace continue\par
\par\smallskip\noindent\textbf{S5. Benchmark suite; the spruce} (from t+62.5\,h)\par
\noindent{\ttfamily t+62.5\,h}\enspace Get the full spruce genome and add it to the benchmark. Put everything required to run the benchmark in one folder, and create a simple script for me to run it reproducibly. Store results in text files (or json maybe) as "\textless{}some machine identifier\textgreater{}-timestamp". Each run = one file.\par
\noindent{\ttfamily t+62.5\,h}\enspace The benchmark should include not only Pinho and mawmetal, but also any other code publsihed in between from the other publications that we can download an run\par
\noindent{\ttfamily t+62.7\,h}$^q$\enspace Benchmark should have 2 version --- one full one to run here, and another that can run on a Macbook retina 2012 with 16GB of ram. A CLI switch to select is good :)\par
\noindent{\ttfamily t+68.0\,h}\enspace continue\par
\par\smallskip\noindent\textbf{S6. Larger genomes; download scripts} (from t+68.0\,h)\par
\noindent{\ttfamily t+68.0\,h}\enspace Could our code process \url{https://www.cell.com/iscience/fulltext/S2589-0042(24)01111-8} ? That's larger than the Spruce... If not, what needs to be changed ? If yes, add it to the benchmark. Do we already have a script that rebuilds the data folder with genomes automatically if it gets deleted ? Can we have a script that also downloads the code of others ? We can't distribute the data and the code of others, but if just have a files with links and a script, it makes reproducibility easy even without redistributing the code.\par
\par\smallskip\noindent\textbf{S7. The fern genome: limits} (from t+68.8\,h)\par
\noindent{\ttfamily t+68.8\,h}\enspace Any chances to run the genome of New Caledonian fork fern (Tmesipteris oblanceolata) with mawmetal ? If not, what would be the limiting factor ? (don't do anything - just analyze)\par
\par\smallskip\noindent\textbf{S8. Project notes for the agent (/init)} (from t+70.8\,h)\par
\noindent{\ttfamily t+70.8\,h}\enspace /init\par
\par\smallskip\noindent\textbf{S9. List-mode memory; large-genome tuning} (from t+71.2\,h)\par
\noindent{\ttfamily t+71.2\,h}\enspace The list mode causes quite a bit more memory usage than the list-lest mode for mawmetal. Can't we optimize and reduce that ?\par
\noindent{\ttfamily t+71.9\,h}\enspace Are there any further optimization in speed or memory that you can see ? Be thorough in your search. If you see speed-memory tradeoff, present the tradeoff and let me decide.\par
\noindent{\ttfamily t+72.3\,h}\enspace Do 1 to 5. Definitely. A. Check again after\par
\noindent{\ttfamily t+72.3\,h}\enspace Do 1 to 5. Definitely. A. Check again after \#4 as suggested B. 270Mb isn't much. Keep the fast one C. Check again after \#4\par
\noindent{\ttfamily t+72.4\,h}$^q$\enspace in the benchmark, allow swap growth by 24GB before killing\par
\noindent{\ttfamily t+72.4\,h}$^q$\enspace add mit license to all the new code that we wrote\par
\noindent{\ttfamily t+73.0\,h}\enspace Can you add the full human genome to the table above ?\par
\noindent{\ttfamily t+73.1\,h}$^q$\enspace well.... We mostly need optimizations for large genomes! That's where it matters. Can you review what we've done above, and the assumptions that we used, focusing our effort on large genomes ? (human, spruce)\par
\noindent{\ttfamily t+73.4\,h}\enspace check on the spruce genome too!\par
\noindent{\ttfamily t+85.4\,h}\enspace Do 1, 5 and 6 For 2,3,4 we made lots of changes to libsais in another discussion. Check the changes and see if the proposals are still relevant.\par
\noindent{\ttfamily t+86.2\,h}\enspace do 3 then, and that's it\par
\par\smallskip\noindent\textbf{S10. libsais, libcubwt; tail and deep stage} (from t+73.6\,h)\par
\noindent{\ttfamily t+73.6\,h}\enspace Look at libsais. \url{https://github.com/IlyaGrebnov/libsais} First, the github page links to something even faster - \url{https://github.com/IlyaGrebnov/libcubwt} . Is it relevant for us ? Can we further optimize either of those libraries in a generic case ? Can we make them faster specifically on Apple silicon by using metal ? Don't do anything now, just analyze\par
\noindent{\ttfamily t+73.9\,h}\enspace Go on, do 1 and 2. Then we'll decide for 3 and 4\par
\noindent{\ttfamily t+74.3\,h}\enspace so what's the conclusion ? What's best we can do here ?\par
\noindent{\ttfamily t+74.3\,h}\enspace yes, go ahead\par
\noindent{\ttfamily t+74.4\,h}$^q$\enspace You also said "Making the tail scan read the PLCP array through the suffix array, instead of keeping a separate LCP array, would save 5.2 GB." What's the drawback of doing this ?\par
\noindent{\ttfamily t+75.4\,h}\enspace Do A, B and C. Before that - where are you doing the changes ? On the third\_party/libsais or elsewhere ?\par

}

\end{article}

\end{document}